\documentclass[aps,floatfix,prd,twocolumn,preprintnumbers,superscriptaddress,groupedaddress]{revtex4}  
\usepackage{amsmath,graphicx,epsfig,amssymb,multirow}
\RequirePackage{mathptmx,flushend}
\usepackage[utf8]{inputenc} 
\usepackage{graphicx}  
\usepackage[usenames,dvipsnames]{xcolor}
\usepackage{dcolumn}   
\usepackage{bm}        
\usepackage{amssymb}   
\usepackage{multirow}
\usepackage{slashed,cancel}
\usepackage{hyperref}
\hypersetup{colorlinks, citecolor=ForestGreen, linkcolor=blue, urlcolor=blue}
\usepackage{subfiles}
\usepackage{calrsfs}
\usepackage[a4paper, vmargin=2.0cm, hmargin=2.0cm, binding offset=0cm]{geometry}

\begin{document}

\preprint{HRI-RECAPP-2023-01}
\title{Measuring Electroweak Quantum Numbers of Color Sextet Resonances at the LHC}
\author{Soubhik~Kumar}
\email{soubhik@berkeley.edu}
\affiliation{Berkeley Center for Theoretical Physics, Department of Physics, University of California,
	Berkeley, CA 94720, USA}
\affiliation{Theoretical Physics Group, Lawrence Berkeley National Laboratory, Berkeley, CA 94720,
	USA}
\author{Rafiqul~Rahaman}
\email{rafiqulrahaman@hri.res.in}
\affiliation{Regional Centre for Accelerator-based Particle Physics, Harish-Chandra Research Institute, A CI of Homi Bhabha National Institute, 
	Chhatnag Road, Jhunsi, Prayagraj, 211019, India}

\author{Ritesh~K.~Singh}
\email{ritesh.singh@iiserkol.ac.in}
\affiliation{Department of Physical Sciences, 
	Indian Institute of Science Education and Research Kolkata,  
	Mohanpur, 741246, India}

\begin{abstract}
We study the prospect of measuring the electroweak quantum numbers of beyond the Standard Model (SM) color sextet particles that decay into same-sign top quark pairs.
Among these particles, the color sextet scalars give rise to top quarks with the same chirality, while the top quarks coming from the color sextet vector would have opposite chirality.
This difference gets encoded in the angular distributions of the bottom quarks and leptons originating from the decays of the top quarks.
We utilize this feature and the energy distributions of the final state jets and leptons to distinguish among the three possible color sextet resonances, taking into account various SM background processes at the $13$ TeV LHC.
\end{abstract}

\maketitle

\section{Introduction}\label{sec:intro} 
A variety of beyond the Standard Model (BSM) scenarios, especially those addressing the Higgs hierarchy problem, e.g., supersymmetry or composite Higgs, predict new physics around the TeV scale (see Ref.~\cite{ParticleDataGroup:2022pth} for reviews). 
The search for such BSM states has been actively going on at the Large Hadron Collider (LHC) and will continue through its high-luminosity (HL-LHC) phase, in conjunction with other indirect probes.

Given the absence of new physics at the LHC so far, we can ask a bottom-up and purely group theoretic question as follows.
Noting that the SM is based on the gauge group $SU(3)_c\times SU(2)_L\times U(1)_Y$, we can ask which possible BSM scalar or vector particles can have direct, tree-level couplings to SM fermions.
A priori, there are a number of such BSM states~\cite{Ma:1998pi, DelNobile:2009st}.
However, a subset of them would couple to both leptons and quarks so as to mediate proton decay at tree level, unless that is forbidden by some other global symmetry, and therefore are ruled out~\cite{Arnold:2012sd} for TeV-scale masses.
Among the remaining states, those coupling to two top quarks are particularly interesting from an experimental perspective.
As we will discuss in the following, two such scalar and one such vector resonance have the SM quantum numbers, 
\begin{equation}\label{eq:q_no}
\Phi_1 \sim ({\bf{6}}, {\bf{1}}, 4/3), ~\\
\Phi_3 \sim ({\bf{6}}, {\bf{3}}, 1/3),~ \\
\Phi_2^\mu \sim ({\bf{6}}, {\bf{2}}, 5/6).
\end{equation}
At the LHC, these states can be produced via their couplings to the first-generation quarks, and subsequently, they can decay into a \textit{like}-sign top quark pair.
The top quarks can then decay into a pair of $b$-jets, a pair of like-sign leptons, and neutrinos when the intermediate $W$ bosons decay leptonically.
The phenomenology and ultraviolet origin of such color sextet diquarks have been discussed extensively in the literature, see, e.g.,~\cite{Ma:1998pi, Chacko:1998td, Cakir:2005iw, Mohapatra:2007af, Chen:2008hh, Bauer:2009cc, DelNobile:2009st, Han:2009ya, Arnold:2009ay, Zhang:2010kr, Han:2010rf, Berger:2010fy, Giudice:2011ak, Arnold:2012sd, Chivukula:2015zma}.

The question we would like to ask in the present work is the following: suppose a discovery of a color sextet particle is made at the (HL-)LHC.
Then purely based on the lab-frame observables, can we extract the quantum numbers of the discovered sextet state and distinguish among the three possible sets of quantum numbers of a generic color sextet particle, as mentioned in Eq.~(\ref{eq:q_no})?

To answer this, we first note that the fermionic couplings of interest are inherently chiral in nature.
Therefore, the top quarks coming from the sextet decay would carry definite polarization and this plays a crucial role in extracting the quantum numbers of the sextet states~\cite{Berger:2010fy,Zhang:2010kr}.
For example, the rest-frame angular distributions of leptons and $b$-jets are different depending on the polarization of the parent top quarks, as we describe explicitly below. 
However, since in such processes, the top quarks decay leptonically, there is missing energy in the form of neutrinos in the final state, and reconstructing the rest frame of a top quark is not immediate.
While it is possible to use the $M_{T2}$ observable~\cite{Lester:1999tx, Barr:2003rg} to address this issue~\cite{Cho:2008tj,Guadagnoli:2013xia}, it is still useful to construct observables that do not rely on any such rest frame reconstruction.
To this end, we construct some new lab-frame observables that can be used to investigate and isolate the quantum numbers of the various sextet states.

Some lab-frame observables for processes involving missing energy have been discussed in the literature. One example is the visible energy fraction of the leptons~\cite{Shelton:2008nq,Berger:2012an}
\begin{equation}\label{eq:energy-frac}
z_i = \frac{E_{l_i}}{E_{l_i}+E_{b_i}}.
\end{equation} 
Here $E_{l_i}$ and $E_{b_i}$ are respectively the energies of the lepton and the associated $b$-quark originating from the decay of the same top quark. 
This observable is sensitive to the polarization of the top quark. 
However, in the signal of our interest, we have two $b$-jets and two leptons, and the effectiveness of $z_i$ inherently depends on correctly pairing a $b$-quark with the associated lepton. 
This motivates us to look for additional lab-frame observables which are independent of such pairing. 
We will show that such observables can be constructed based on the azimuthal distribution of final state visible particles. 

The rest of this work is organized as follows. 
In Sec.~\ref{sec:model}, we describe the couplings of the sextet states to the SM fermions and the existing experimental constraints on such states.
In Sec.~\ref{sec:obs}, we construct the lab-frame observables of interest and explain how top quark polarization and spin correlation plays an important role in this context through a parton-level analysis at the $13$ TeV LHC.
In Sec.~\ref{sec:result}, we present our results to distinguish among the color sextet states through a detector-level simulation, taking into account possible SM backgrounds.
We conclude in Sec.~\ref{sec:conclusion}.

\section{Model}\label{sec:model}
In this work, we focus on color sextet BSM resonances that couple to two top quarks.
Given the quantum numbers of the SM quark doublet $q_L = \left(\begin{tabular}{c}
	$u_{L}$ \\ 
	$d_{L}$ \\
\end{tabular}
\right)
\sim ({\bf{3}}, {\bf{2}}, +1/6)$, and the right-handed up-type quark $u_{R} \sim ({\bf{3}}, {\bf{1}}, +2/3)$, the quantum numbers of the BSM scalar and vector resonances are fixed as in~(\ref{eq:q_no}).
We first consider the two scalars, $\Phi_1$ and $\Phi_3$.
We can write the Yukawa coupling of $\Phi_1$ as (see, e.g.,~\cite{Han:2009ya})
\begin{equation}\label{singletcoupling}
\lambda_{1}\bar{K}^a_{ij}\Phi_{1a} \bar{u}_{Ri}  u_{Rj}^c + {\rm h.c.},
\end{equation}
where the superscript $c$ denotes charge conjugation operation. The matrices $\bar{K}^a_{ij}$ are determined by the Clebsch-Gordon coefficients for the sextet representation of $SU(3)$. In both Eq.~(\ref{singletcoupling}) and Eq.~(\ref{tripletcoupling}) below, the indices $a$ and $i,j$ correspond to color indices and they run over $1\cdots 6$ and $1\cdots 3$, respectively. 

The Yukawa coupling of $\Phi_3$ is given in an analogous manner, except that it couples to the symmetric combination of two copies of $q_L$. The coupling to $u_{L}$ is given by
\begin{equation}\label{tripletcoupling}
\lambda_{3}\bar{K}^a_{ij}\Phi_{3a} \bar{u}_{Li} u_{Lj}^c + {\rm h.c.}.
\end{equation}
Here we have focused on the part of the isospin triplet field that couples only to the up-type quarks since that can decay into a pair of top quarks, which is the signal of our interest.
Both in~(\ref{singletcoupling}) and~(\ref{tripletcoupling}), we have suppressed the generation indices on the Yukawa couplings, $\lambda_1$ and $\lambda_3$.

Finally, the coupling of the vector resonance is described by,
\begin{equation}\label{vectorcoupling}
\lambda_{2}\bar{K}_{ij}^a\Phi_{2,a}^\mu \bar{u}_{Rj}\gamma_\mu u_{Li}^c +  {\rm h.c.}. 
\end{equation}
Here we have focused on the up-quark coupling for the same reason as above.
We will be agnostic about how $\Phi_2^\mu$ gets its mass.

For a general Yukawa coupling, there would be large flavor changing neutral current processes mediated by the above sextet states.
To avoid the stringent experimental constraints from those, we assume that the matrices $\lambda_1$, $\lambda_2$, and $\lambda_3$ are flavor diagonal.

With this assumption, the next stringent constraints on $\lambda_1$ come from the measurements of $D^{0}-\bar{D}^{0}$ mixing \cite{Chen:2009xjb},  
\begin{equation}
\frac{|\text{Re}(\lambda_{1,cc}\lambda_{1,uu}^*)|}{m_\phi^2} < x \times 7.2\times 10^{-11} / \text{GeV}^{2},
\end{equation}
where $x = \Delta m_D/\Gamma_D $
is a $D^0-\bar{D}^0$ mixing parameter. 
Assuming $CP$ conservation and taking $x\approx 4\times 10^{-3}$ \cite{LHCb:2021ykz}, we get
$\text{Re}(\lambda_{1,cc}\lambda_{1,uu}^*) < 3\times 10^{-7}$ for $m_\phi= 1~ \text{TeV}$. 
In the following, we will assume a hierarchy between $\lambda_{1,cc}$ and $\lambda_{1,uu}$ with $|\lambda_{1,cc}| \ll |\lambda_{1,uu}|$ so as to satisfy the above bound.
The dominant constraints on $\lambda_2$ and $\lambda_3$ also come from $D^0-\bar{D}^0$ mixing~\cite{Zhang:2010kr}, which can again be suppressed by taking the couplings to second generation quarks to be small as we mentioned above.

We note that $\Phi_1$ and $\Phi_3$ decay to a pair of right-handed and left-handed top quarks, respectively.
This implies that we would be able to distinguish between the quantum numbers of these two sextet scalars using the polarization properties of the  top quarks.
In particular, in its rest frame, a right-handed top quark would decay into leptons whose average distribution would peak in the same direction as the top quark spin, while the associated $b$ quark distribution would be peaked in the opposite direction.
Thus after boosting to the lab frame, the angle between the two leptons, each coming from the two daughter top quarks from $\Phi_1$ decay, will be peaked around $\Delta\phi =\pi$, while the angle between the two $b$ quarks will be more broadly distributed around $\Delta\phi =\pi$.
The situation with $\Phi_3$ is exactly the opposite of this.
While the vector BSM resonance $\Phi_2^\mu$ decays into a same-sign top pair as well, the top quarks would have opposite chirality.
These features can then be used to distinguish among $\Phi_1$, $\Phi_2^\mu$, and $\Phi_3$, as we will see below.
We will refer to these particles as {\tt Singlet}, {\tt Doublet}, and {\tt Triplet}, respectively, based on their $SU(2)_L$ quantum numbers.
For numerical simulations, we choose the benchmark: $m_{\phi} = 1$~TeV, $\lambda_{uu} =\lambda_{tt} = 0.003$ for all the three sextet particles. 
We note that there is an upper bound of $1.2$ pb on the $tt$ production cross-section given by CMS   coming  from the same-sign di-lepton, missing energy, and jets search~\cite{CMS:2017tec}. However, the 
$tt$ production cross-sections in our models are much smaller compared to the CMS limit. For example, $\sigma_{tt}\simeq0.02$ pb for the Singlet case in our chosen benchmark, which is only 1\% of the CMS limit.

\section{Observable}\label{sec:obs}
\begin{figure*}
	\centering
	\includegraphics[width=0.48\textwidth]{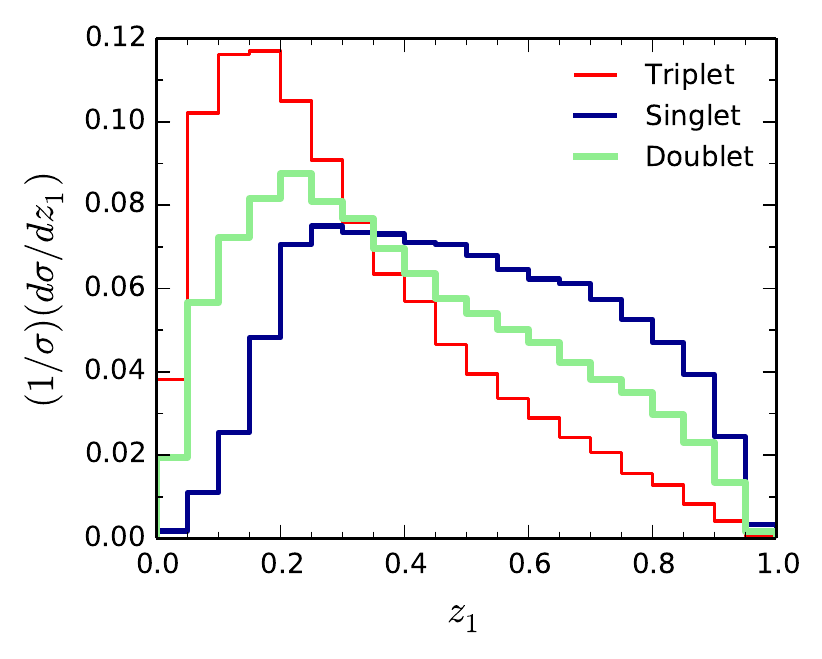}
	\includegraphics[width=0.48\textwidth]{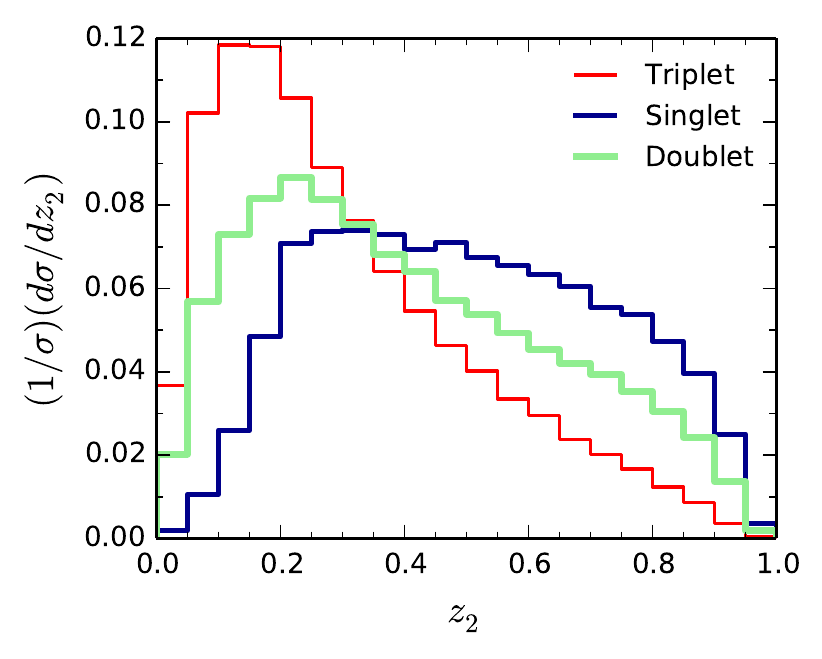}	
	\caption{Normalized distributions for the visible energy fractions  $z_1$ and $z_2$ for the signals with parton level  events. See text for further discussions.
	}
	\label{fig:parton_1d}
\end{figure*}
\begin{figure*}
	\centering
	\includegraphics[width=0.3\textwidth]{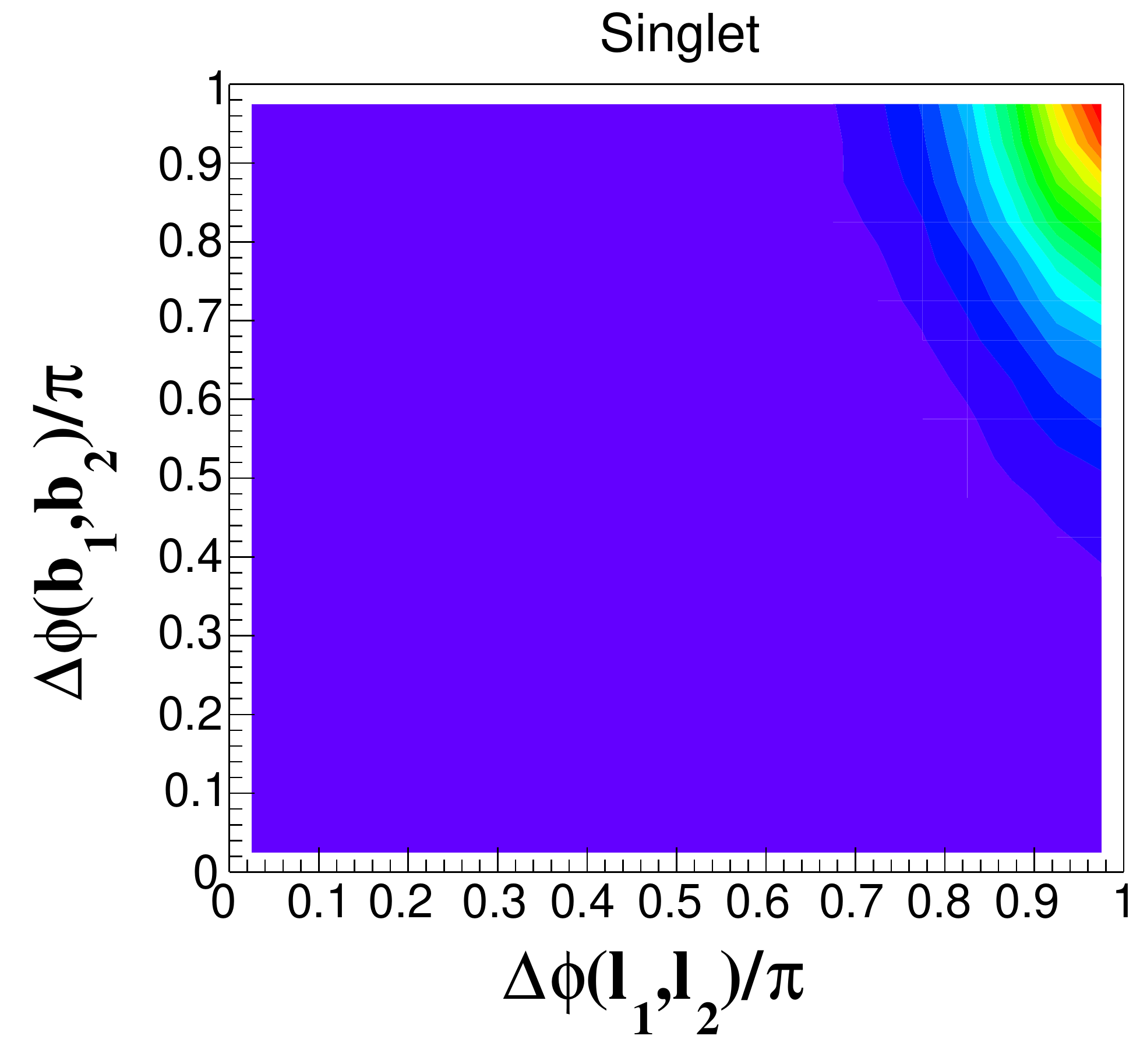}	
	\includegraphics[width=0.3\textwidth]{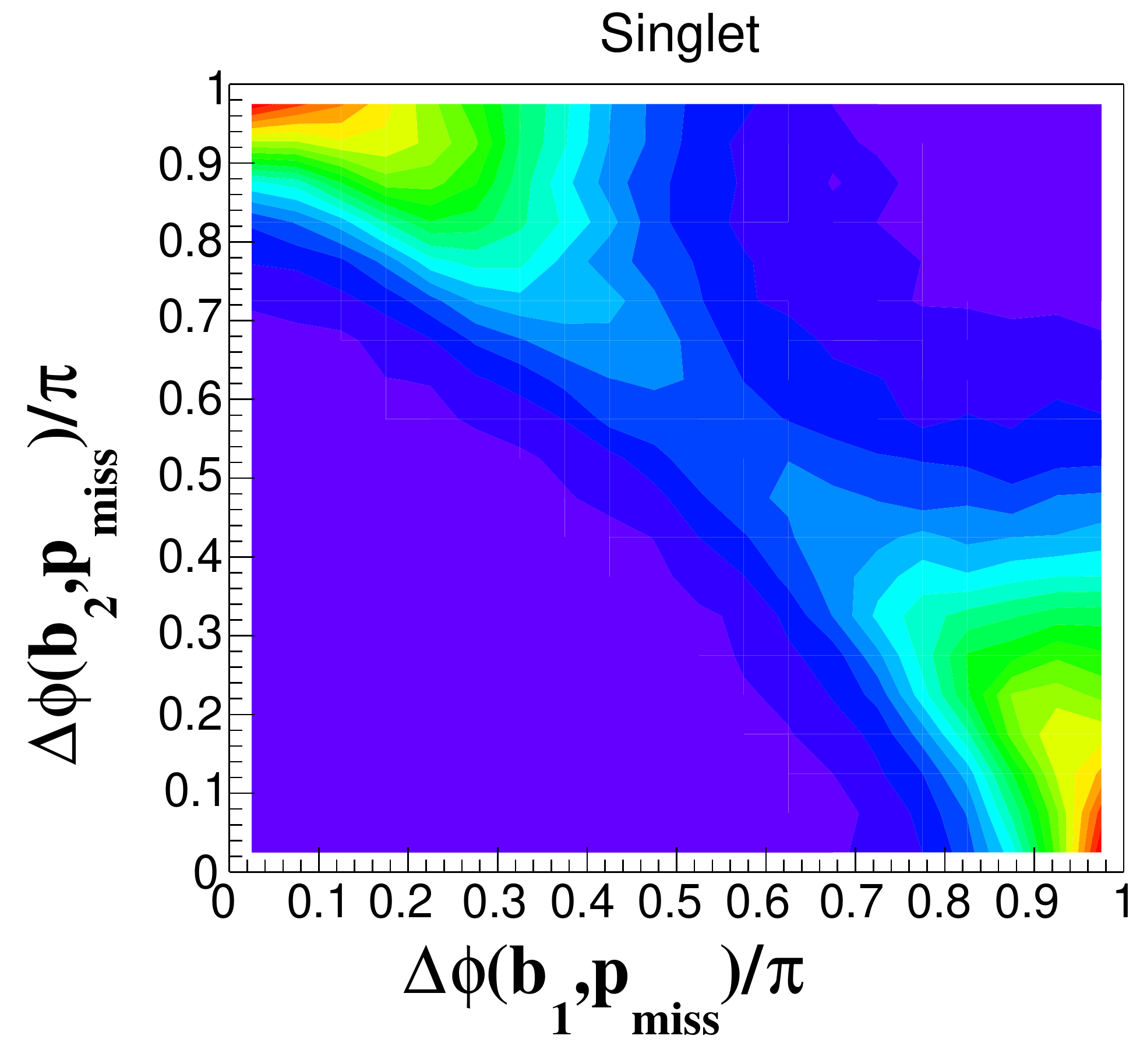}	
	\includegraphics[width=0.3\textwidth]{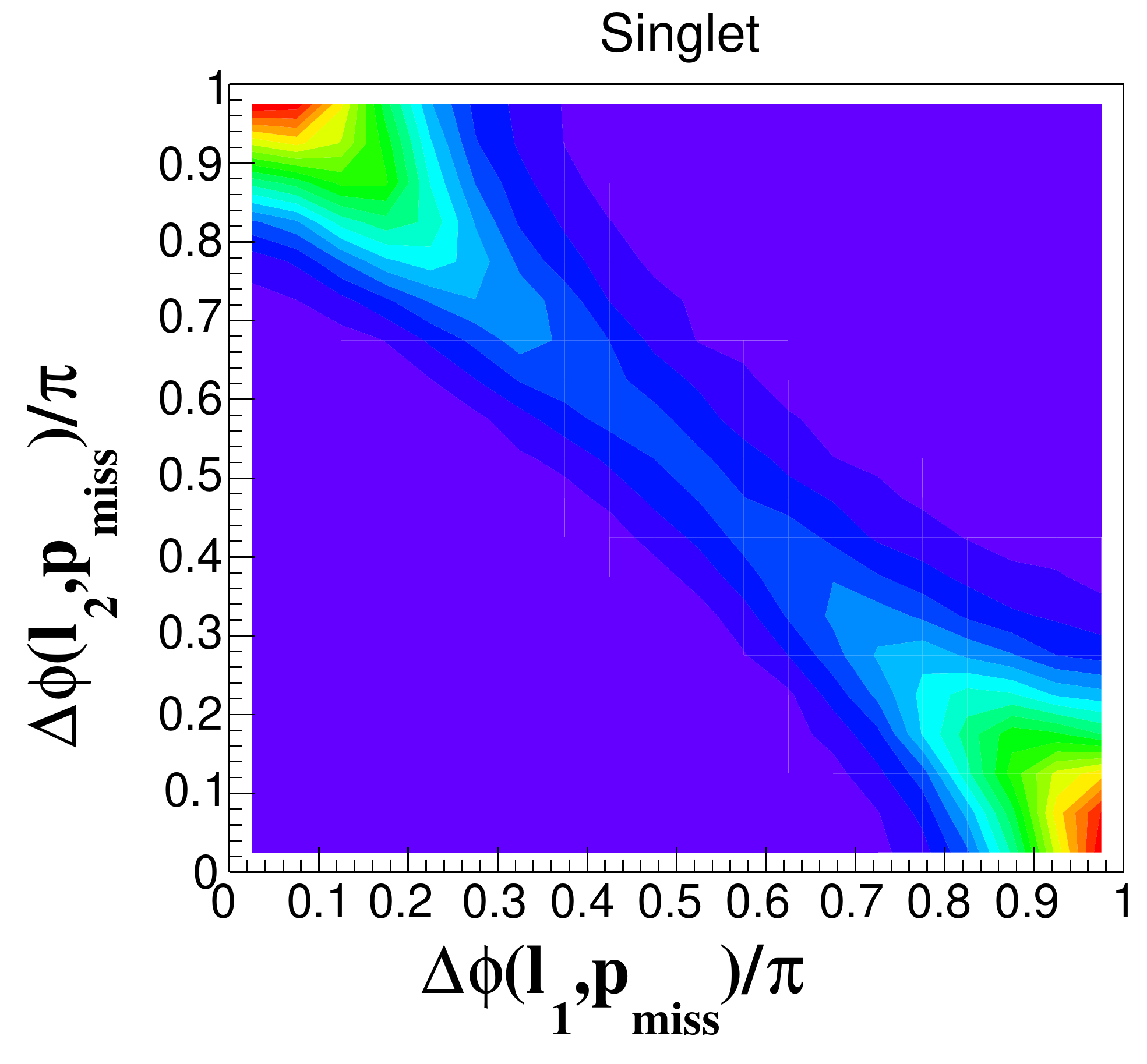}	
	\includegraphics[width=0.3\textwidth]{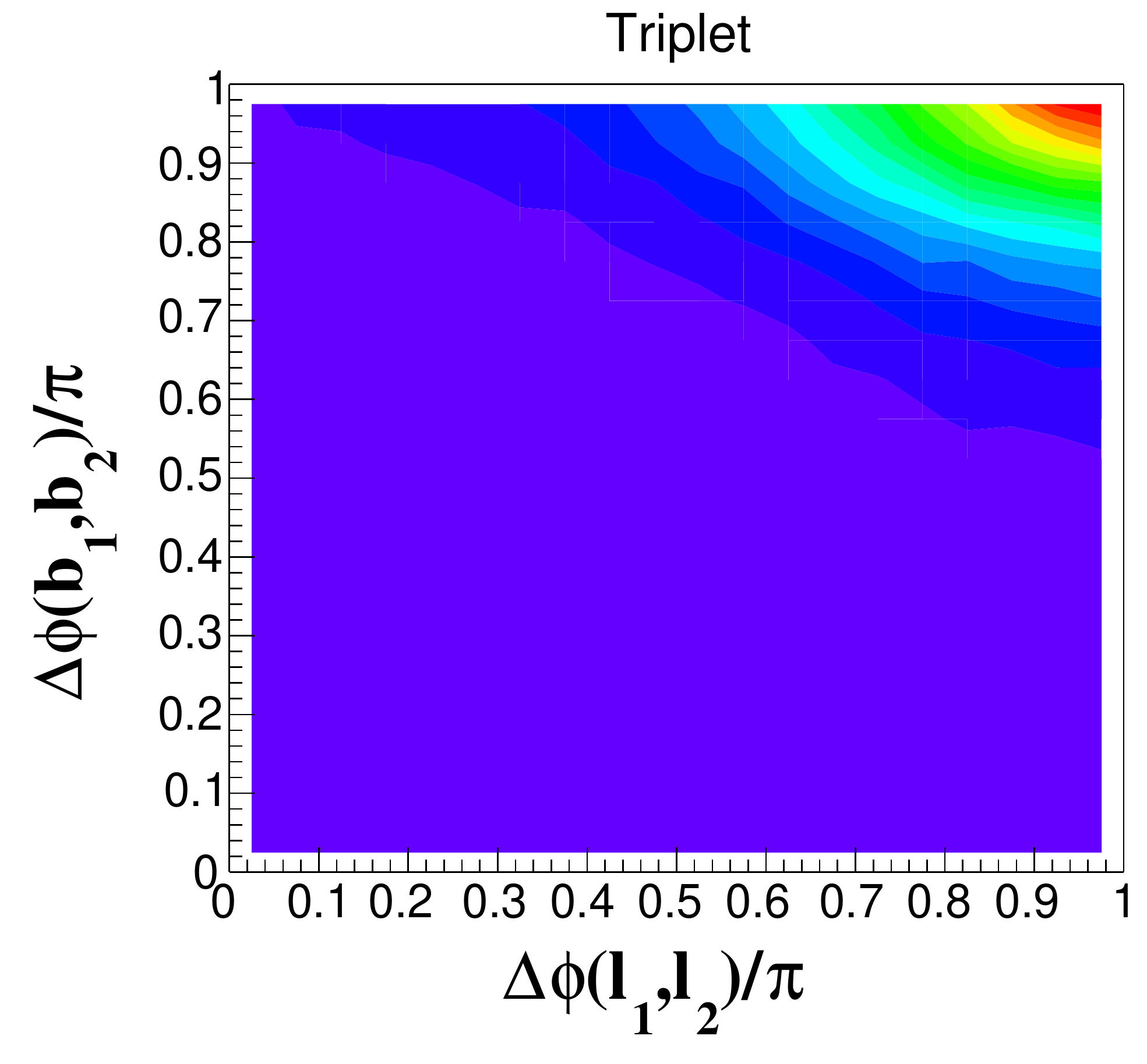}	
	\includegraphics[width=0.3\textwidth]{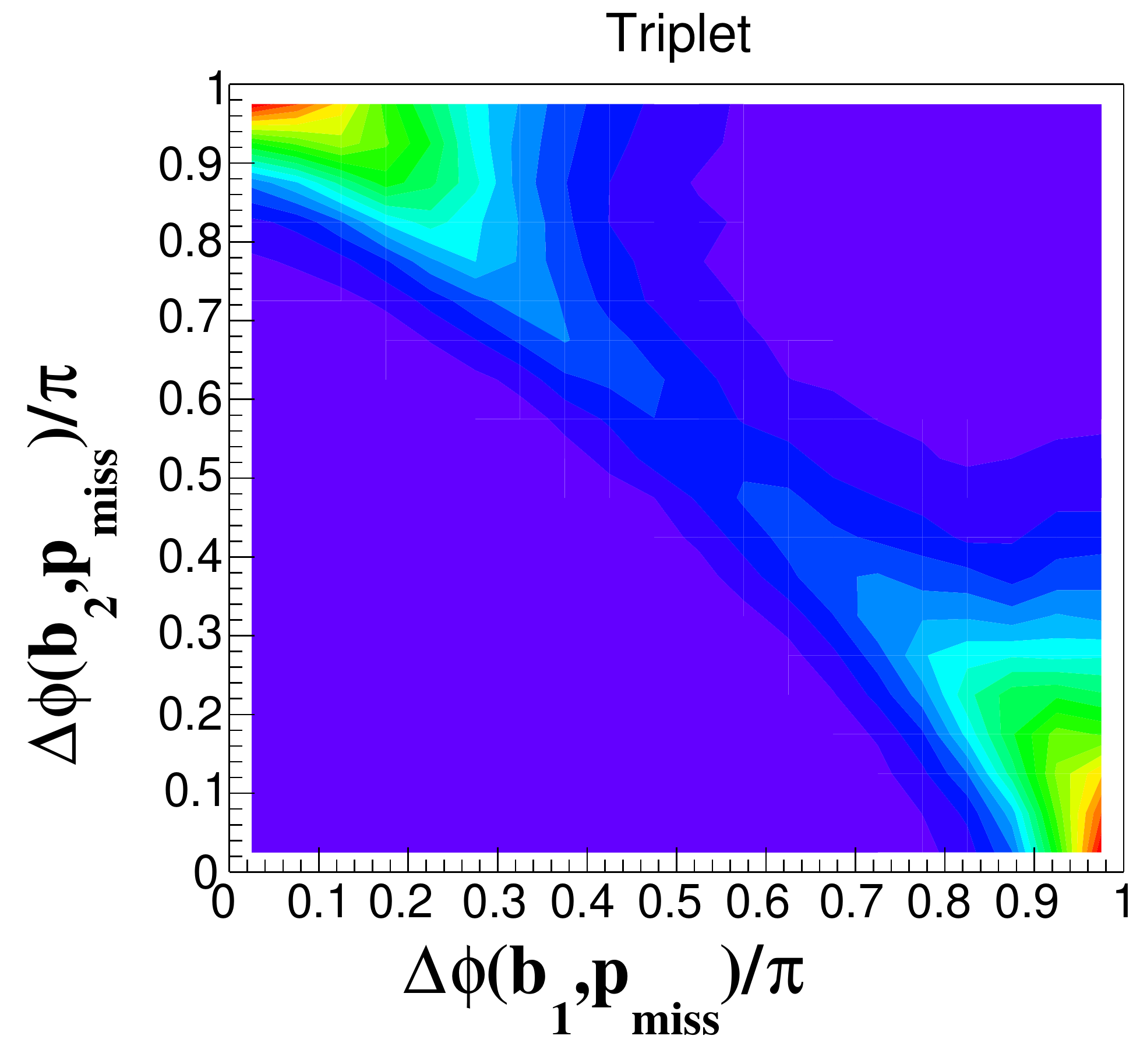}	
	\includegraphics[width=0.3\textwidth]{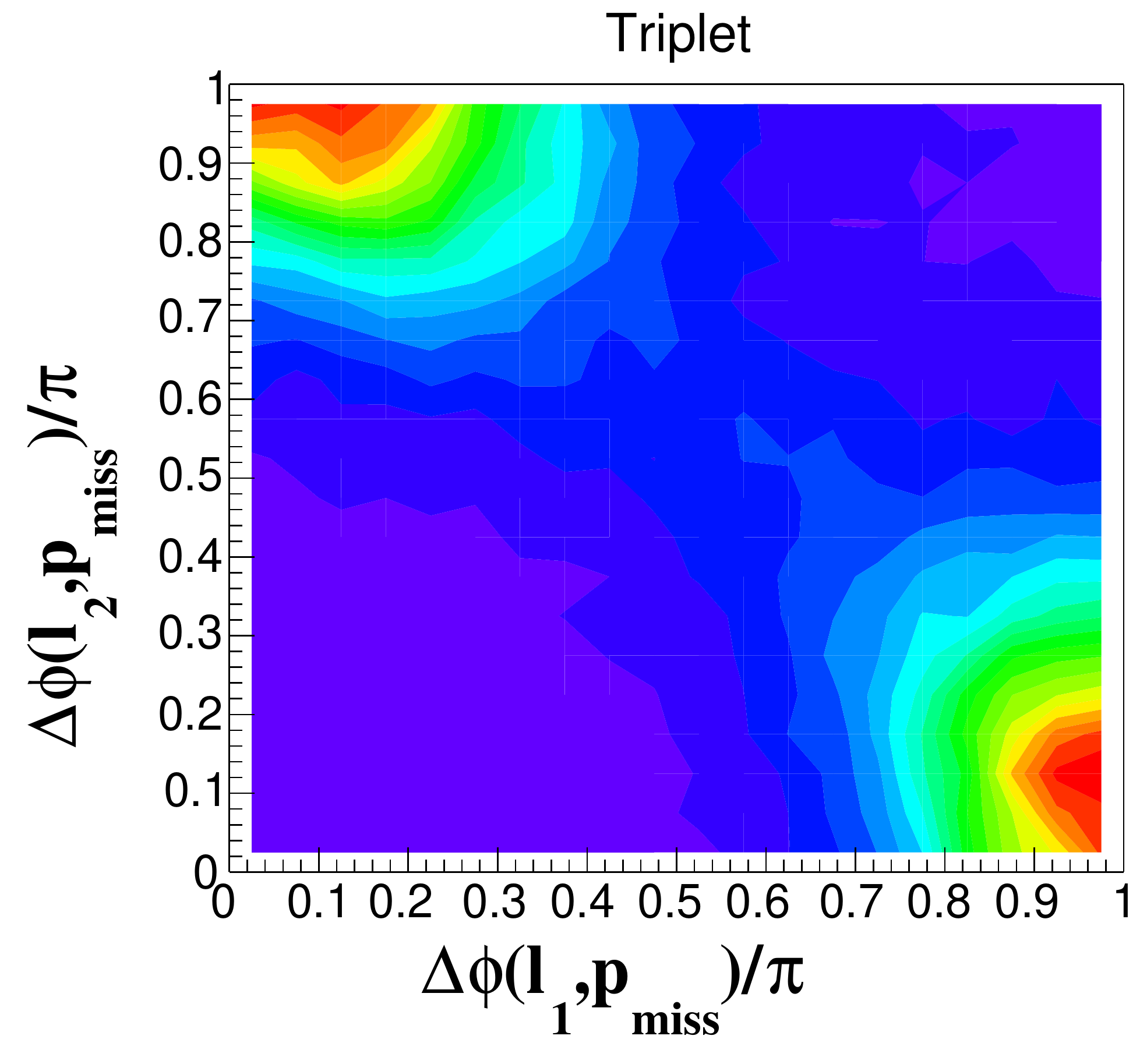}	
	\includegraphics[width=0.3\textwidth]{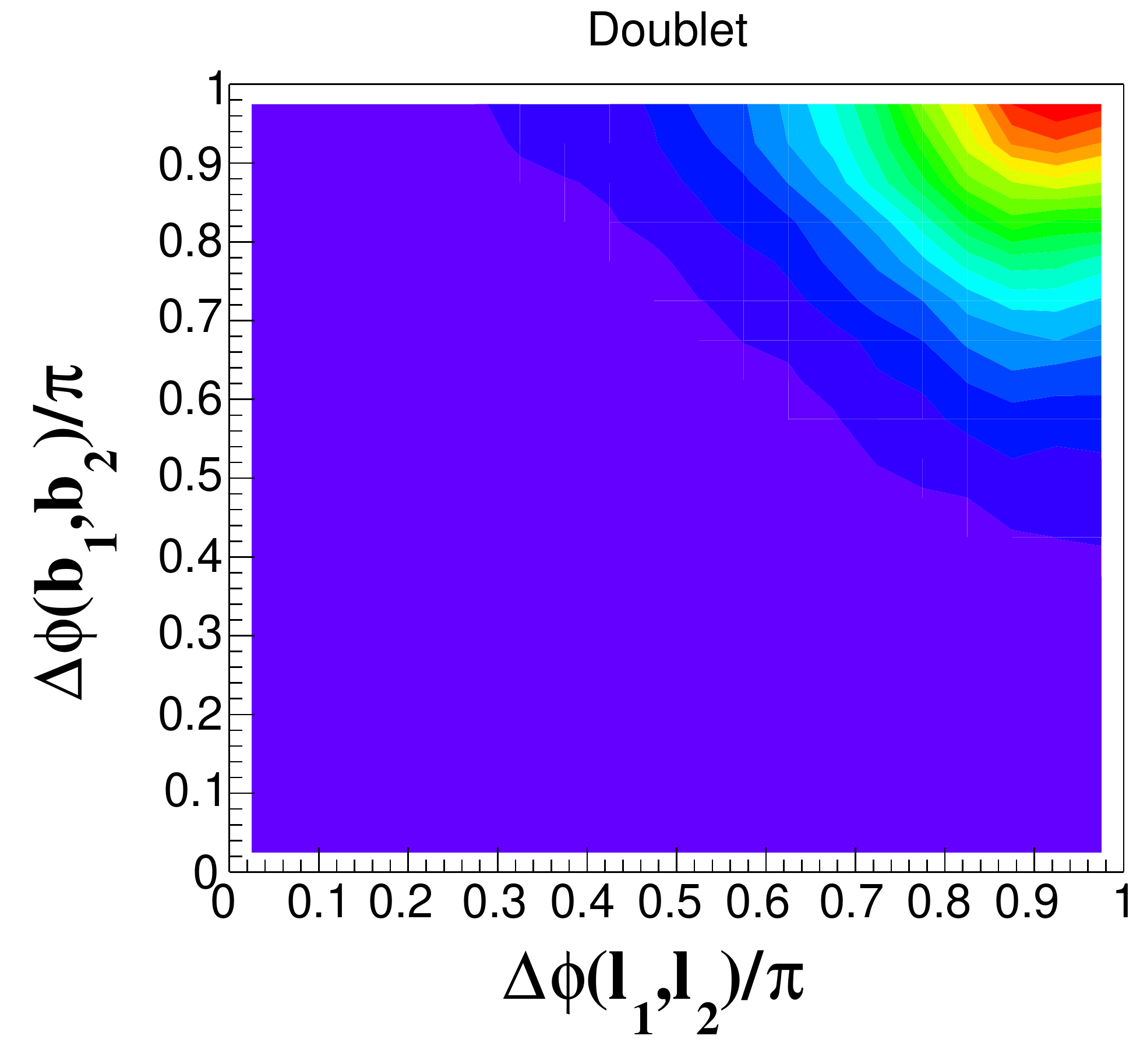}	
	\includegraphics[width=0.3\textwidth]{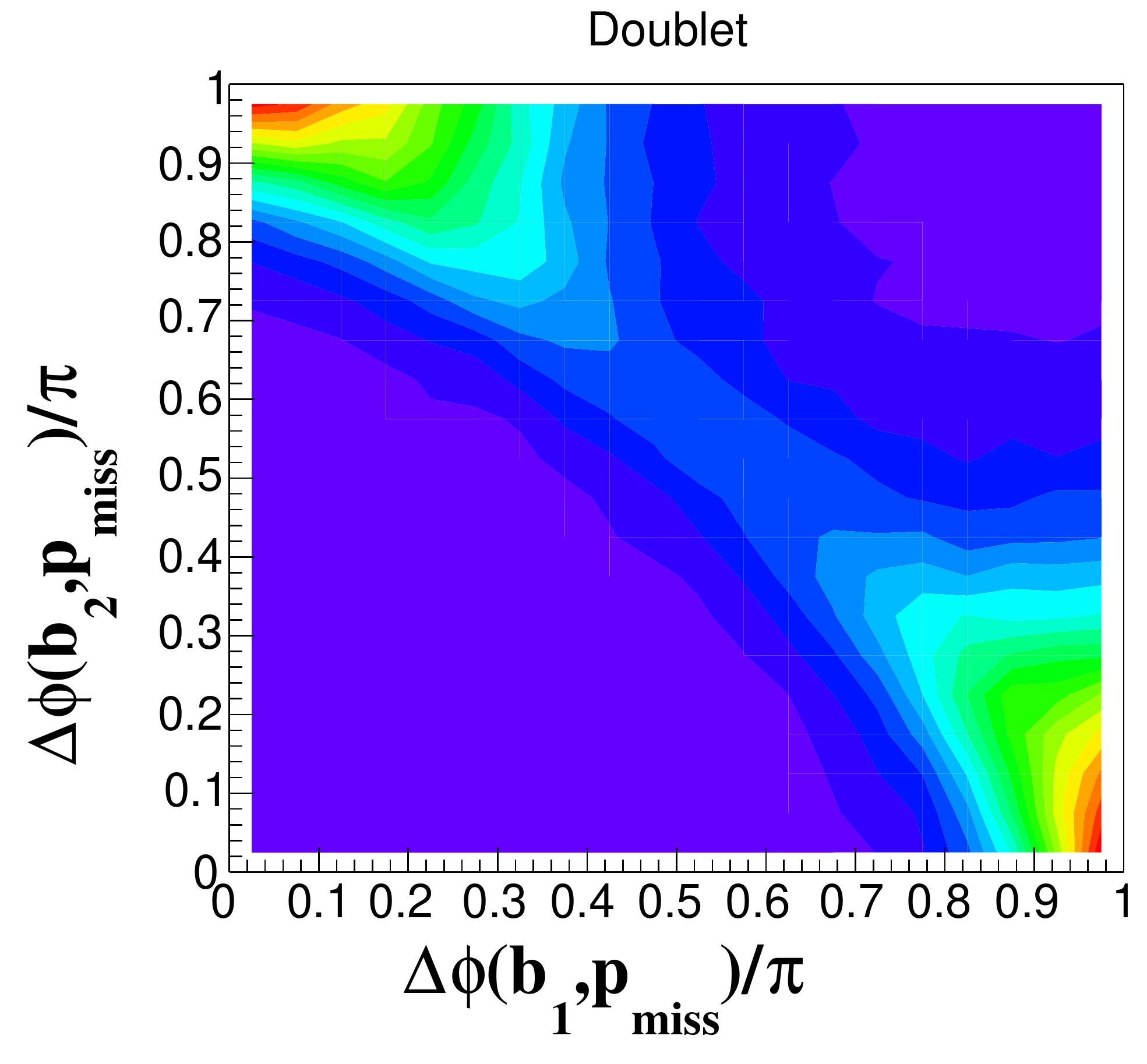}	
	\includegraphics[width=0.3\textwidth]{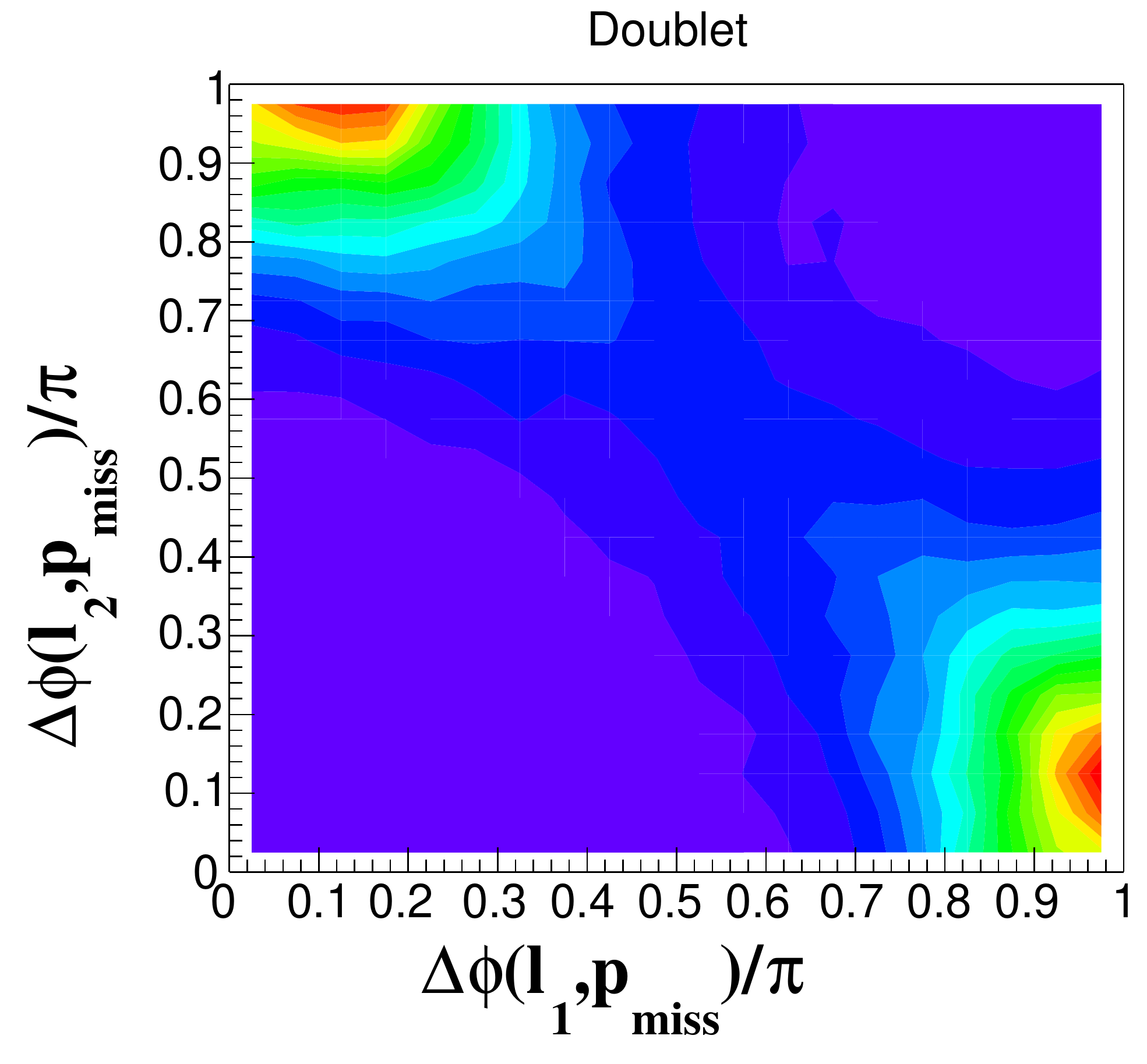}	
	\caption{Normalized distributions for the two-dimensional correlations  between $\Delta\phi(l_1,l_2)$ vs. $\Delta\phi(b_1,b_2)$ (left-column), $\Delta\phi(b_1,p_{\text{miss}})$ vs. $\Delta\phi(b_2,p_{\text{miss}})$ (middle-column), and $\Delta\phi(l_1,p_{\text{miss}})$ vs. $\Delta\phi(l_2,p_{\text{miss}})$ (right-column) for the signals with parton level events.
		The red points represent maximum density, and light blue represents minimum density for the number of events. 
	}
	\label{fig:parton_2d}
\end{figure*}
The polarization of a top quark is preserved in its decay products and this plays an important role in our analysis.
The double angular distribution of daughter fermions from the two top quarks can be expressed as~\cite{Bernreuther:1993df} 
\begin{eqnarray}\label{angular}
	&&\frac{1}{\sigma}\dfrac{d^2\sigma}{(d\cos\theta_1)(d\cos\theta_2)} =  \nonumber \\
	&&\dfrac{1}{4\pi}\left(1+\alpha_1 p_{t_1}\cos\theta_1 + \alpha_2 p_{t_2}\cos\theta_2\right.\nonumber \\
&&	\left. 
	+\alpha_1\alpha_2  pp_{tt}\cos\theta_1  \cos\theta_2  \right),
\end{eqnarray}
in terms of the polarization of two top quarks ($p_{t_{1}}, p_{t_{2}}$) and their spin correlation ($pp_{tt}$). Here, $\theta_i$ are the decay angles of the daughter fermions at their respective top quark's rest frame.  
The quantity $\alpha_i$ is called the analyzing power which depends on the daughter fermion. 
For example, for a lepton, $\alpha_l=1$, whereas for a $b$-quark, $\alpha_b=-0.4$~\cite{Boudjema:2009fz}. 
The polarization $p_t$ is positive (negative) for the right (left)-polarized top quark.  
The above formula in Eq.~(\ref{angular}) is valid only in the rest frame of the top quark which requires a complete reconstruction of the missing neutrinos~\cite{CMS:2015rld,CMS:2018adi,CMS:2019nrx,Rahaman:2022dwp}.
The polarization of the top quark can also be obtained from the distribution of the ratio of energies of the daughter $b$-quark $(E_b)$ and the top quark $(E_t)$, $E_b/E_t$.
This again requires a complete reconstruction of the top quark momenta and hence that of the neutrinos. 

The visible energy fractions ($z_i$), as defined in Eq.~(\ref{eq:energy-frac}), are correlated with the polarization of the top quarks, and can let us distinguish between left and right chiral top quarks, without completely reconstructing their momenta.
However, this requires a correct pairing of the two $b$-quarks with the corresponding leptons. 
This can be achieved using the fact that in the lab frame, a sibling lepton and a $b$-quark have a smaller $\Delta R$ than the $\Delta R$ between a cousin lepton and a $b$-quark.
Here, $\Delta R = \sqrt{ \Delta\phi^2 + \Delta\eta^2 }$ where $\phi$ is the azimuthal angle and $\eta$ is the pseudorapidity of a particle. 
We label the two leptons as $l_1$ and $l_2$, ordered according to their transverse momenta ($p_T$), i.e., with $p_T(l_1)>p_T(l_2)$. 
The two $b$-quarks can pair with the leptons in two ways. 
If the correct pairs are such that $l_1$ and $b_1$ come from one top quark, while $l_2$ and $b_2$ come from another top quark, then we expect to have 
$\Delta R(l_1,b_1)+\Delta R(l_2,b_2) < \Delta R(l_1,b_2)+\Delta R(l_2,b_1)$.
This criterion is satisfied for more than $96~\%$ of times for the chosen mass of 1 TeV.
For a heavier scalar, the top quarks will be more boosted which will lead their decay products to collimate further, and this increases the efficiency of pairing the lepton and the $b$-quark correctly.

We study the distributions for the energy fraction variables $z_1$ and $z_2$ defined in Eq.~(\ref{eq:energy-frac}), pairing the leptons and $b$ quarks as described above using parton-level events at $\sqrt{s}=13$ TeV to see how these variables can separate the three signals, namely {\tt Singlet}, {\tt Doublet}, and {\tt Triplet}.  The normalized distributions for $z_1$ and $z_2$  are shown in Fig.~\ref{fig:parton_1d}. 
Both $z_1$ and $z_2$ peak on the lower side for {\tt Triplet} (red lines) as the produced top quarks are left-handed in this case. 
This means the final state leptons are less boosted as we go from the top rest frame to the lab frame, and thus $z_i$'s peak at smaller values. 
For the {\tt Singlet} (blue lines), the variables $z_i$ sharply drop near $0.2$, having more fractional events with $z_i>0.5$, compared to the {\tt Triplet}.
For the {\tt Doublet} (green lines), however, the variables peak below $z_i=0.5$, but the asymmetry with respect to $z_i=0.5$ is smaller than the {\tt Triplet} since the {\tt Doublet} decay gives rise to both left-handed as well as right-handed top quarks.
With realistic detector effects the strengths of these variables might diminish, but the qualitative result would remain the same, as we will see below. 
We define net asymmetries for the visible energy fractions ($z_i$ )  as
\begin{equation}\label{eq:1d_asym}
	{\cal A}_{z_i} = \frac{N(z_i>c_{z_i})-N(z_i<c_{z_i})}{N_{\text{tot}}},
\end{equation}
with $N(x>c)$ being the number of events with $x>c$, and $N_{\text{tot}}$ is the total number of events. 
	We only consider  the $tt$ production process containing $2\ell^+$ final state   and do not include the charge conjugation process  because $\sigma_{\bar{t}\bar{t}}/\sigma_{tt}\approx 0.03$ owing to the small   parton distribution functions (PDFs) at the LHC.

Similar to the polarization parameters, one can construct variables to capture the spin correlation between the two top quarks using the lab frame angular separation between the final state particles in the transverse plane. 
We study two-dimensional (2D) correlations between $\Delta\phi(l_1,l_2)$ vs. $\Delta\phi(b_1,b_2)$, $\Delta\phi(b_1,p_{\text{miss}})$ vs. $\Delta\phi(b_2,p_{\text{miss}})$, and $\Delta\phi(l_1,p_{\text{miss}})$ vs. $\Delta\phi(l_2,p_{\text{miss}})$. 
The normalized 2D correlations are shown in Fig.~\ref{fig:parton_2d} with parton-level events. 
The correlations are not symmetric with respect to the diagonal axes, and they are different for different signals. 
In the first column in Fig.~\ref{fig:parton_2d}, the distributions are more peaked in  $\Delta\phi(l_1,l_2)$ compared to  $\Delta\phi(b_1,b_2)$ for the {\tt Singlet}, while the opposite is true for the {\tt Triplet}.
For the {\tt Doublet}, however, the distributions are less peaked and more symmetric around the $\Delta\phi(l_1,l_2) = \Delta\phi(b_1,b_2)$ line.
This can be explained in the following way.  
As the two top quarks are produced from a heavy resonance, $\Delta\phi(l_1,l_2)$ and $\Delta\phi(b_1,b_2)$ are expected to peak around $\pi$ in the lab frame due to relativistic focusing. 
But due to the positive polarization of top quarks from the {\tt Singlet}, the leptons ($b$-quarks) are emitted primarily in the same (opposite) direction as the top quark spin in the top rest frame.
As a result, after boosting to the lab frame,
the peak of $\Delta\phi(l_1,l_2)$ becomes sharper, while the peak of $\Delta\phi(b_1,b_2)$ becomes broader. 
For the {\tt Triplet} with negatively polarized top quarks, the opposite happens, i.e., the peak of $\Delta\phi(b_1,b_2)$ becomes sharper, while the peak of $\Delta\phi(l_1,l_2)$ becomes broader. 
Finally for the {\tt Doublet}, with a mix of positively and negatively polarized top quarks, the peaks of $\Delta\phi(l_1,l_2)$ and $\Delta\phi(b_1,b_2)$ have similar broadening.

The $\Delta\phi(l/b,p_{\text{miss}})$ peak near zero and $\pi$, as shown in the second and third column of Fig.~\ref{fig:parton_2d}. 
The distribution of $\Delta\phi(l/b,p_{\text{miss}})$ can be explained in a similar way as that of $\Delta\phi(l_1/b_1,l_2/b_2)$, after noting that $\bar{\nu}$ is distributed the same way as $b$ quarks around the top quark spin in the top rest frame.
The peak of $\Delta\phi(l,p_{\text{miss}})$ is sharper than that of $\Delta\phi(b,p_{\text{miss}})$ for the {\tt Singlet}, and the opposite is true for the {\tt Triplet}. For the {\tt Doublet},  $\Delta\phi(l,p_{\text{miss}})$ and $\Delta\phi(b,p_{\text{miss}})$ behave roughly in the same way. 
Given this, we propose the following variables based on these 2D correlations:
\begin{eqnarray}\label{eq:2d_asym}
	x_{lb}&=&\Delta\phi(l_1,l_2)/\pi-\Delta\phi(b_1,b_2)/\pi,\nonumber\\
	x_{ll}&=&\Delta\phi(l_1,p_{\text{miss}})/\pi+\Delta\phi(l_2,p_{\text{miss}})/\pi,\nonumber\\
	x_{bb}&=&\Delta\phi(b_1,p_{\text{miss}})/\pi+\Delta\phi(b_2,p_{\text{miss}})/\pi,
\end{eqnarray}
to extract the spin correlation of the two top quarks.
The asymmetries for the correlations  are defined as,
\begin{eqnarray}\label{aphlb}
	\mathcal{A}_{lb}&=&\frac{N(x_{lb}>0)-N(x_{lb}<0)}{N_{\text{tot}}},\\
\label{aphll}
	\mathcal{A}_{ll} &=& \frac{N(|x_{ll}-1|<c_{ll})-N(|x_{ll}-1|>c_{ll})}{N_{\text{tot}}},\\
\label{aphbb}
\mathcal{A}_{bb} &=& \frac{N(|x_{bb}-1|<c_{bb})-N(|x_{bb}-1|>c_{bb})}{N_{\text{tot}}}.
\end{eqnarray}
Note that these $\Delta\phi$ variables are independent of pairing the leptons and the $b$ quarks, the pairing is required only for the variables $z_i$.
\begin{table}\caption{\label{tab:asymmetris-parton} Values of asymmetries for the signals with parton level events. For all the signals, we choose $\lambda=0.003$ and $m_\phi = 1$ TeV.}
	\begin{tabular*}{0.46\textwidth}{@{\extracolsep{\fill}}lccccc@{}}\hline		
		& $\mathcal{A}_{z_1}$ & $\mathcal{A}_{z_2}$ & $\mathcal{A}_{lb}$ & $\mathcal{A}_{ll}$ & $\mathcal{A}_{bb}$ \\
		\hline
		{\tt Singlet} & $0.31$ & $0.31$  & $0.33$  & $0.58$  & $0.05$  \\ 
		\hline 
		{\tt Triplet} & $-0.36$ & $-0.36$ & $-0.31$  & $0.0$  & $0.28$ \\ 
		\hline 
		{\tt Doublet} & $-0.03$ & $-0.03$  & $-0.10$  & $0.07$ & $0.11$ \\ 
		\hline 
	\end{tabular*}
\end{table}

To maximize the asymmetries defined in Eqs.~(\ref{eq:1d_asym}),  (\ref{aphlb}), (\ref{aphbb}), and (\ref{aphll}), we choose $c_{z_1}=c_{z_2} \simeq 0.38$ and $c_{bb}=c_{ll}\simeq 0.14$ by observing the parton level distributions in Figs.~\ref{fig:parton_1d} and~\ref{fig:parton_2d}. 
The values of the asymmetries at parton level are listed in Table~\ref{tab:asymmetris-parton} for the three signals. 
It is clear that with the magnitudes and signs of the asymmetries, one can easily identify and separate the signals. 
However, we need to extract these asymmetries with realistic detector effects, including possible backgrounds. 
Therefore, in the next section, we analyze the signals with detector-level events in the presence of SM backgrounds.

\section{Results}\label{sec:result}
\begin{figure*}[ht]
	\centering
	\includegraphics[width=0.46\textwidth]{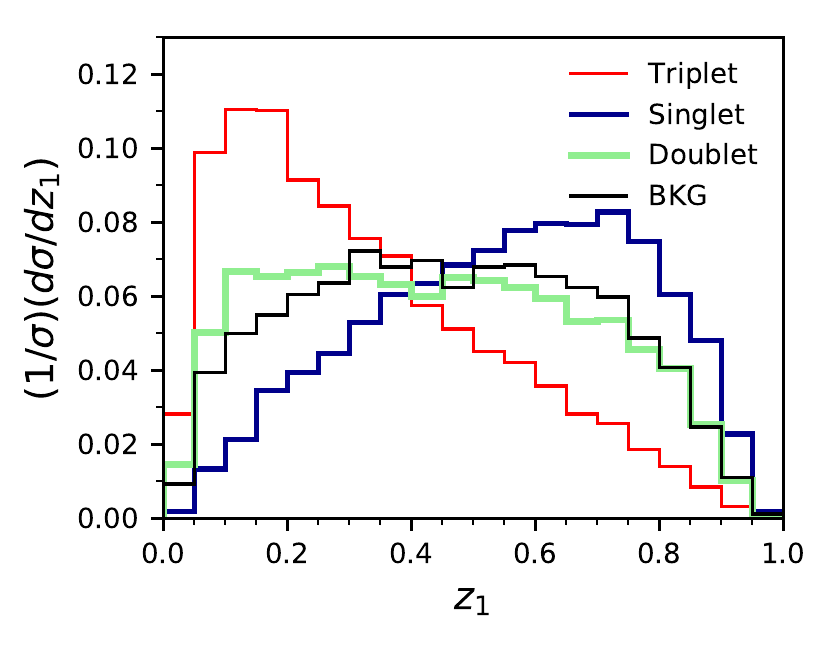}
	\includegraphics[width=0.46\textwidth]{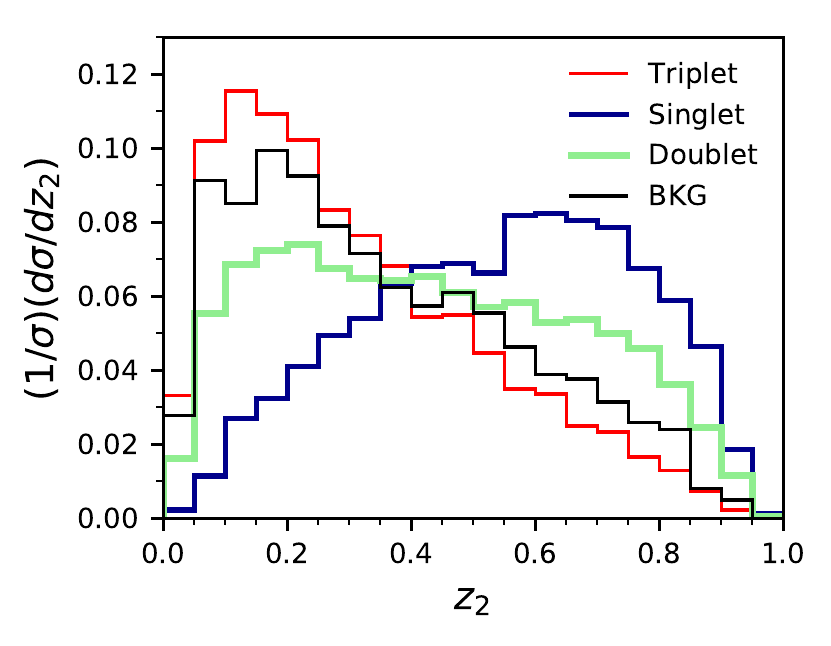}
	\caption{	\label{fig:sig_bkg_hist_1D} Normalized distributions for the visible energy fractions  $z_1$ and $z_2$ for the signals and background at {\tt Delphes} level with selection cuts as in Eq.~(\ref{eq:sel-cut}).}
\end{figure*}
\begin{figure*}[ht]
	\centering
	\includegraphics[width=0.3\textwidth]{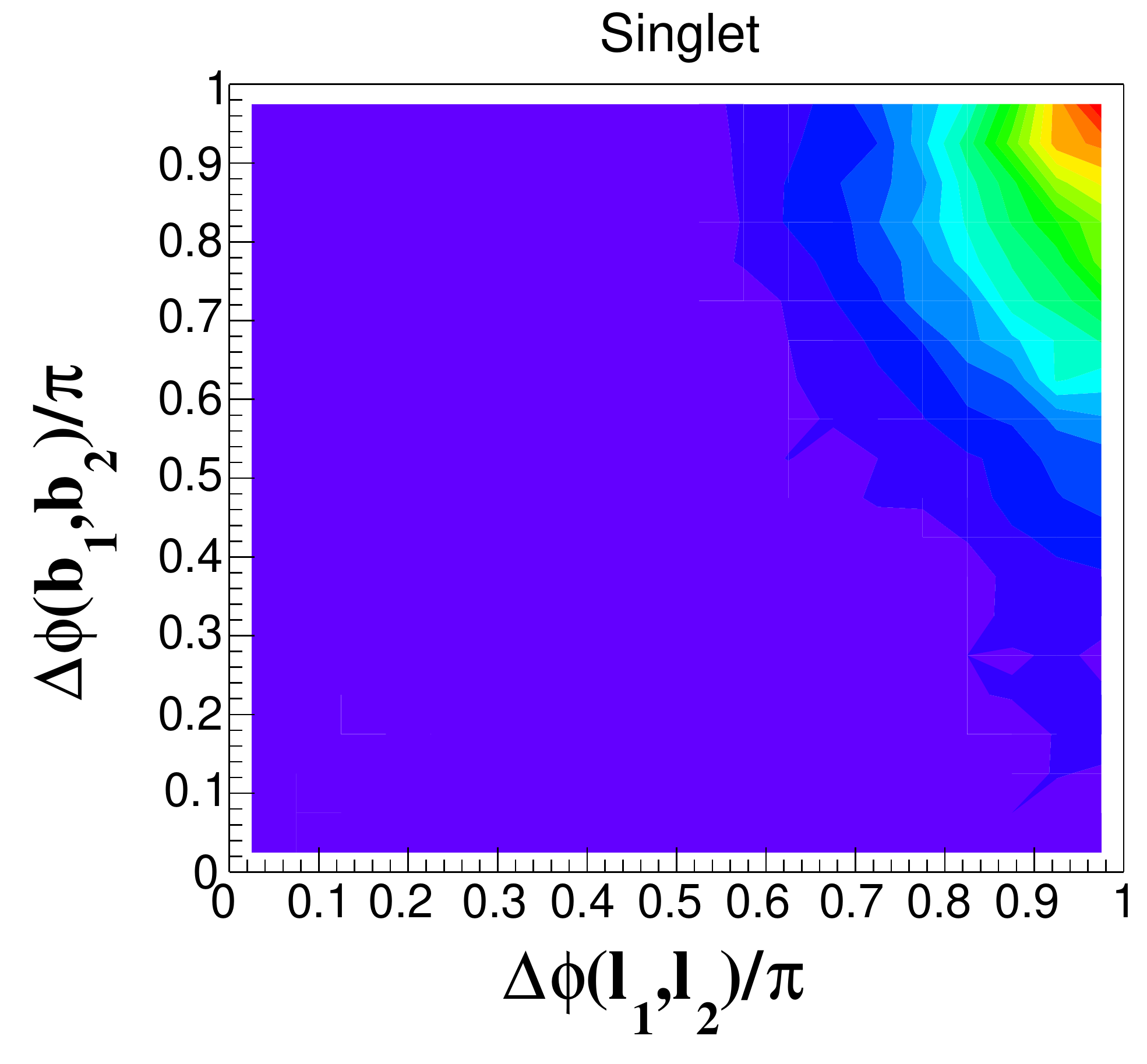}
	\includegraphics[width=0.3\textwidth]{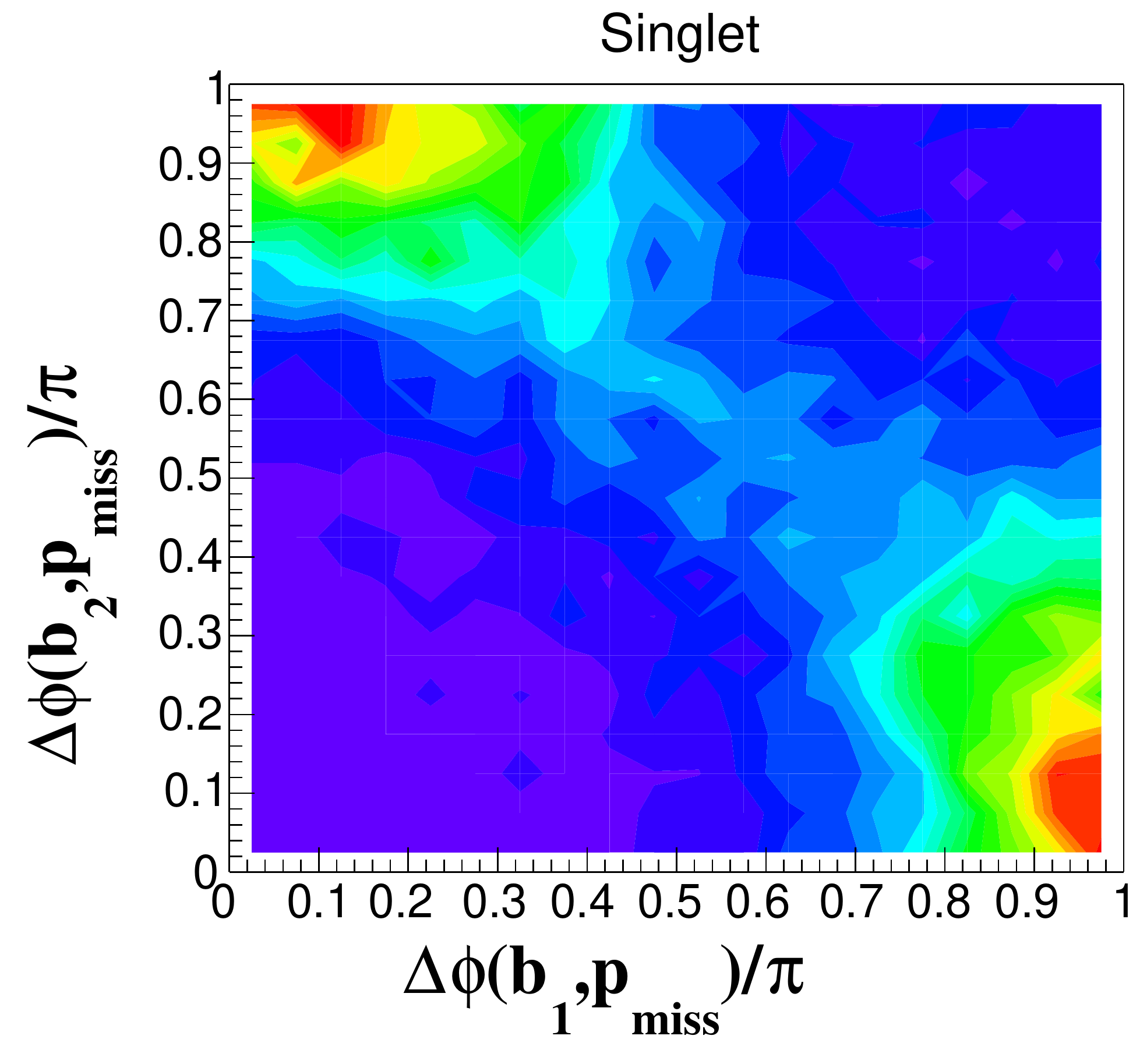}
	\includegraphics[width=0.3\textwidth]{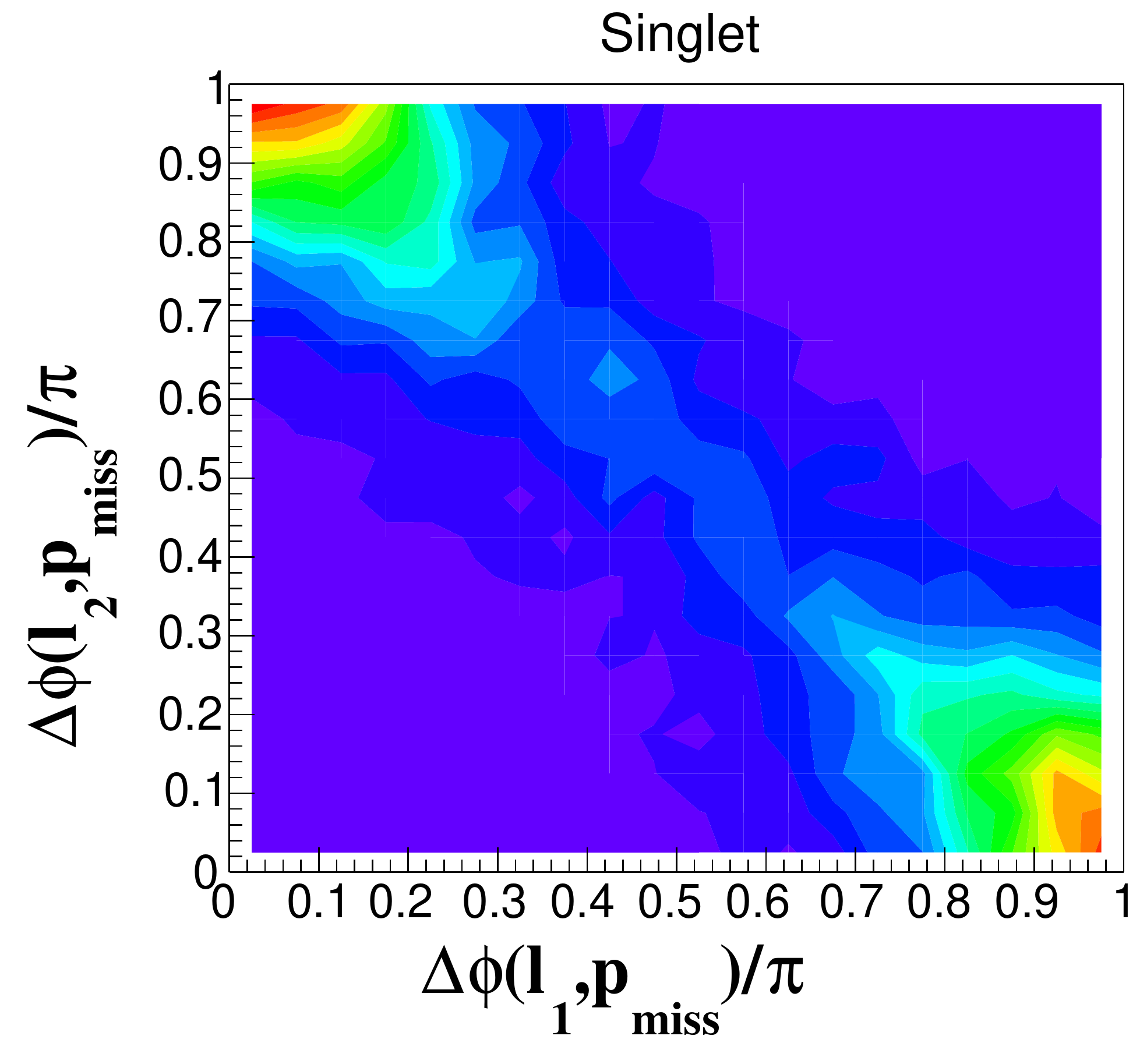}
	\includegraphics[width=0.3\textwidth]{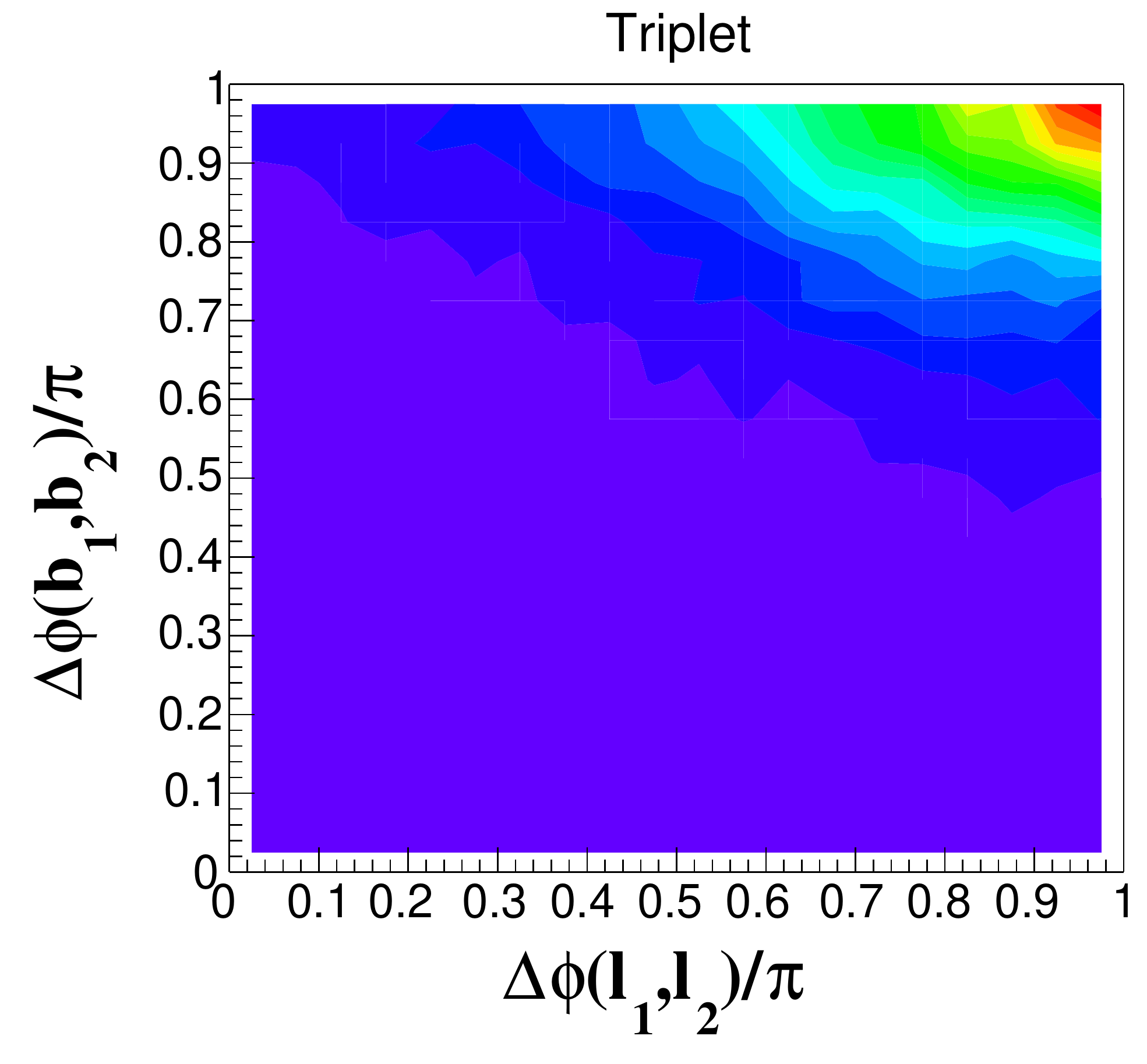}
	\includegraphics[width=0.3\textwidth]{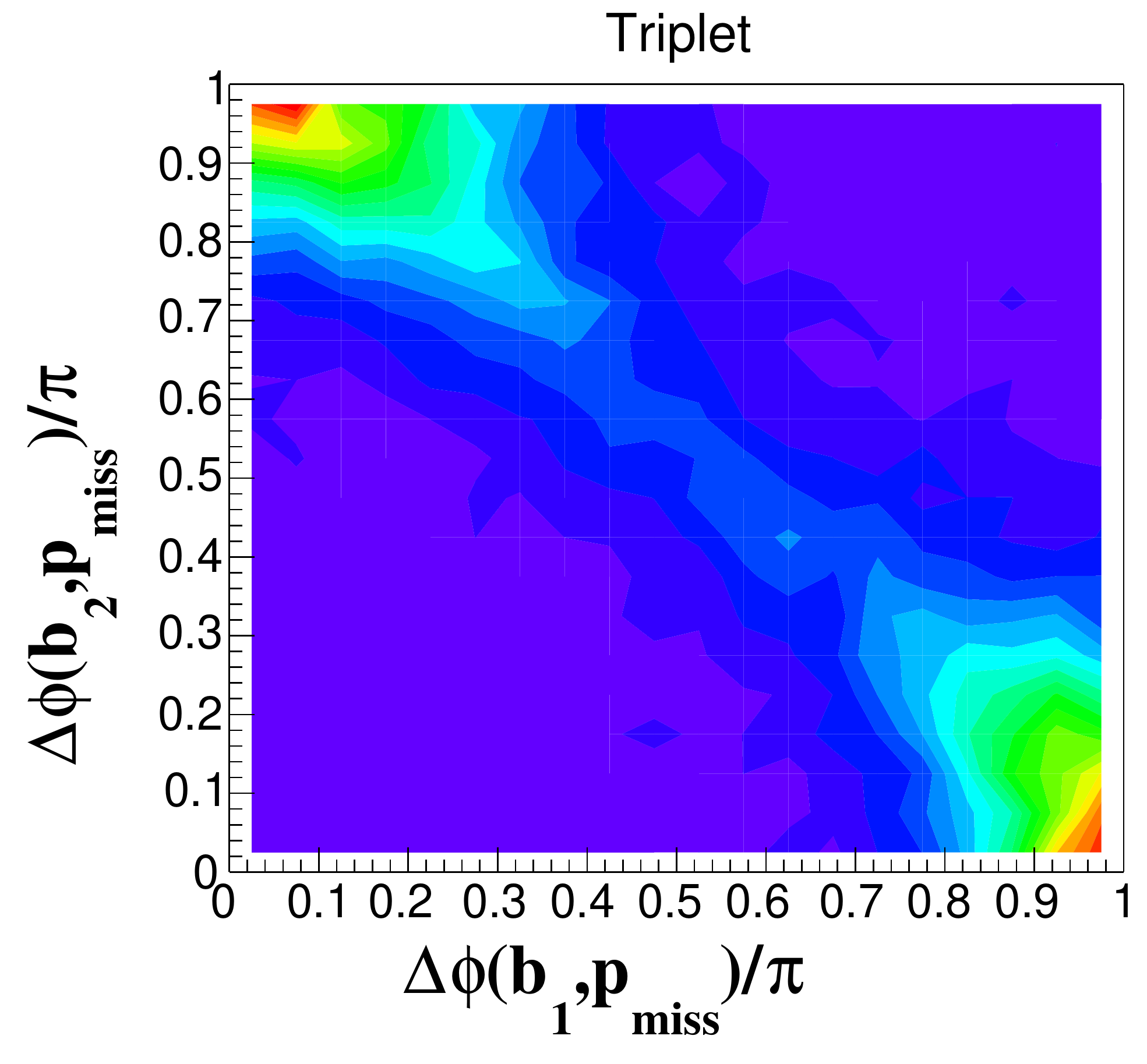}
	\includegraphics[width=0.3\textwidth]{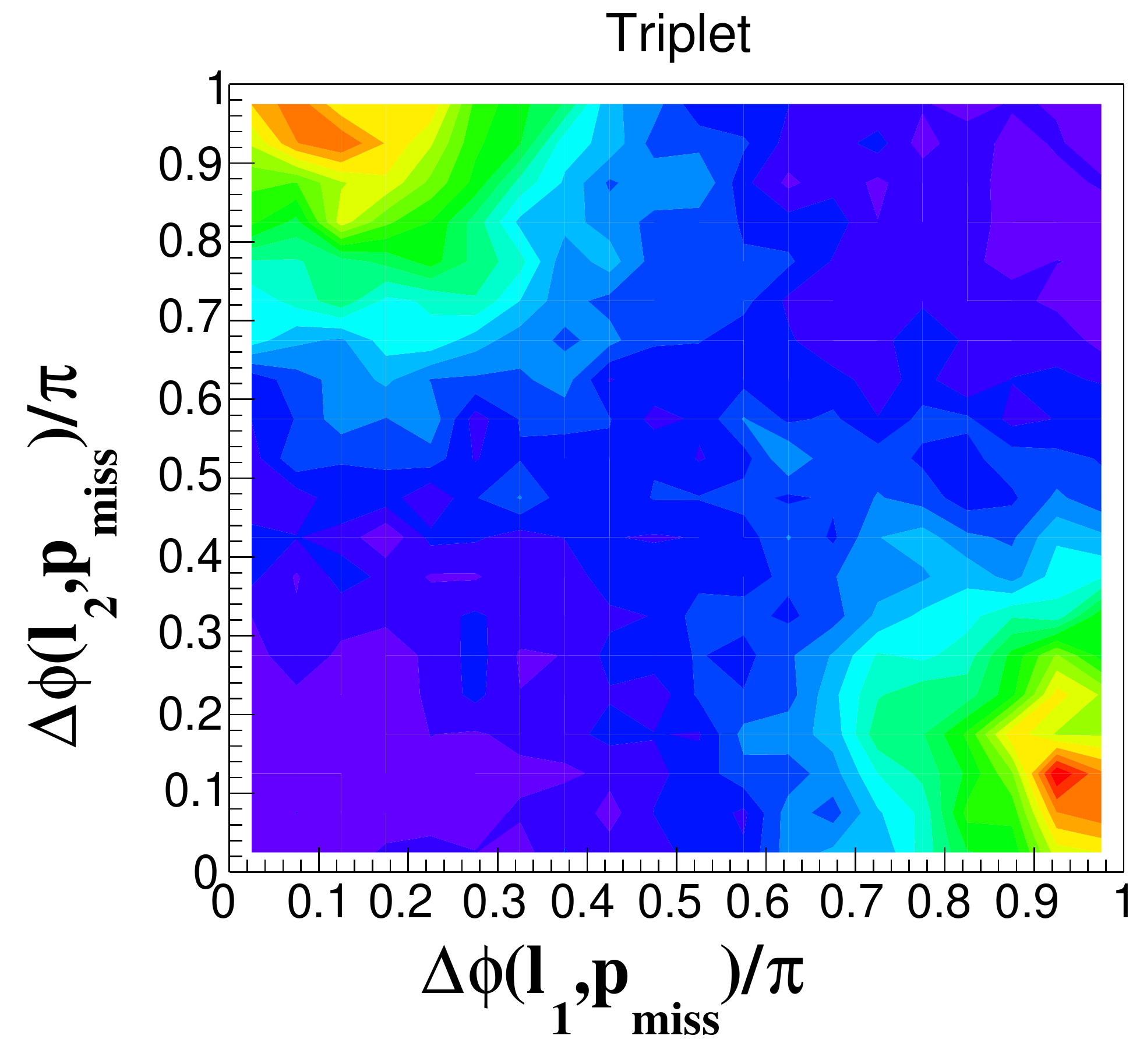}
	\includegraphics[width=0.3\textwidth]{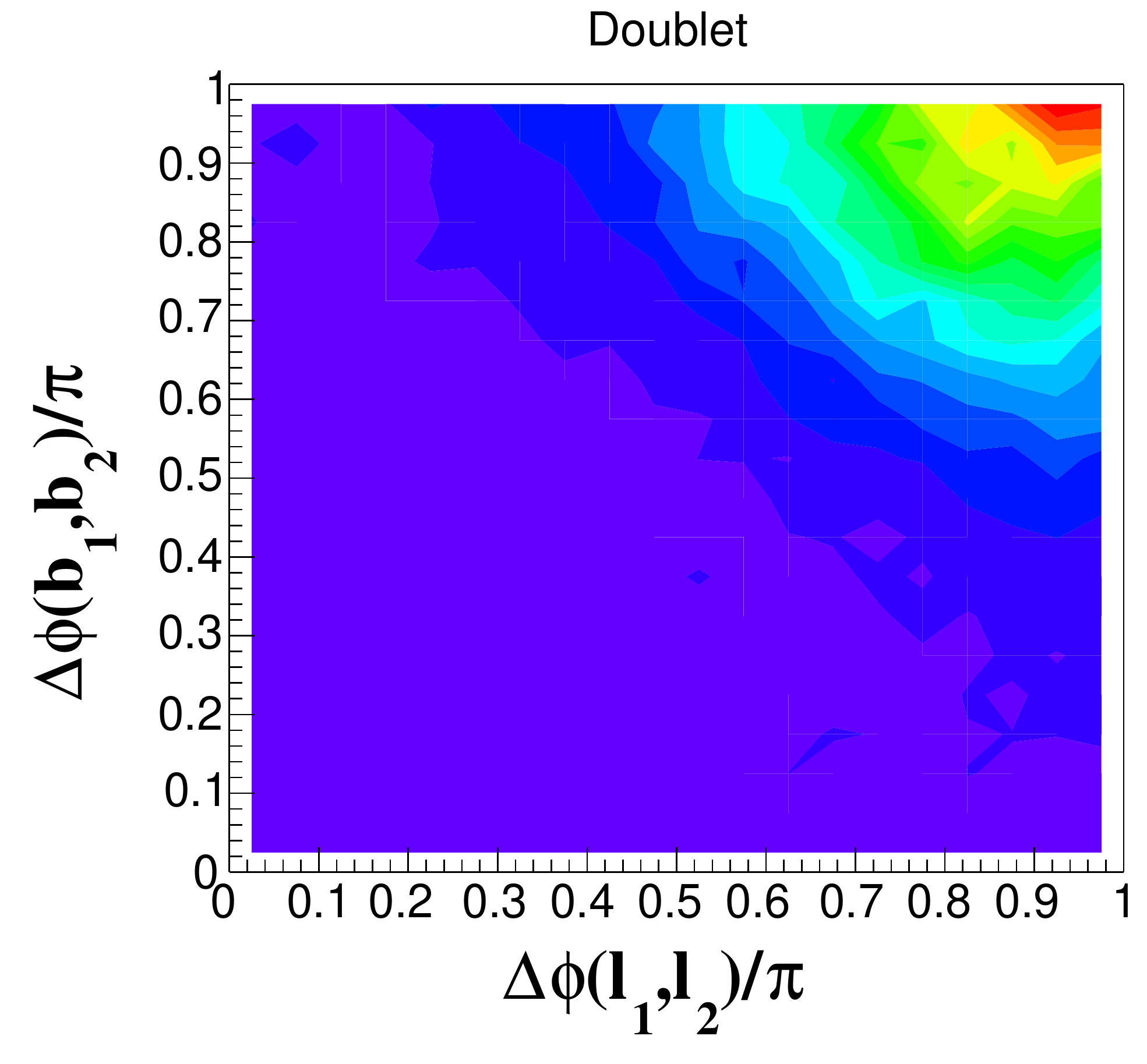}
	\includegraphics[width=0.3\textwidth]{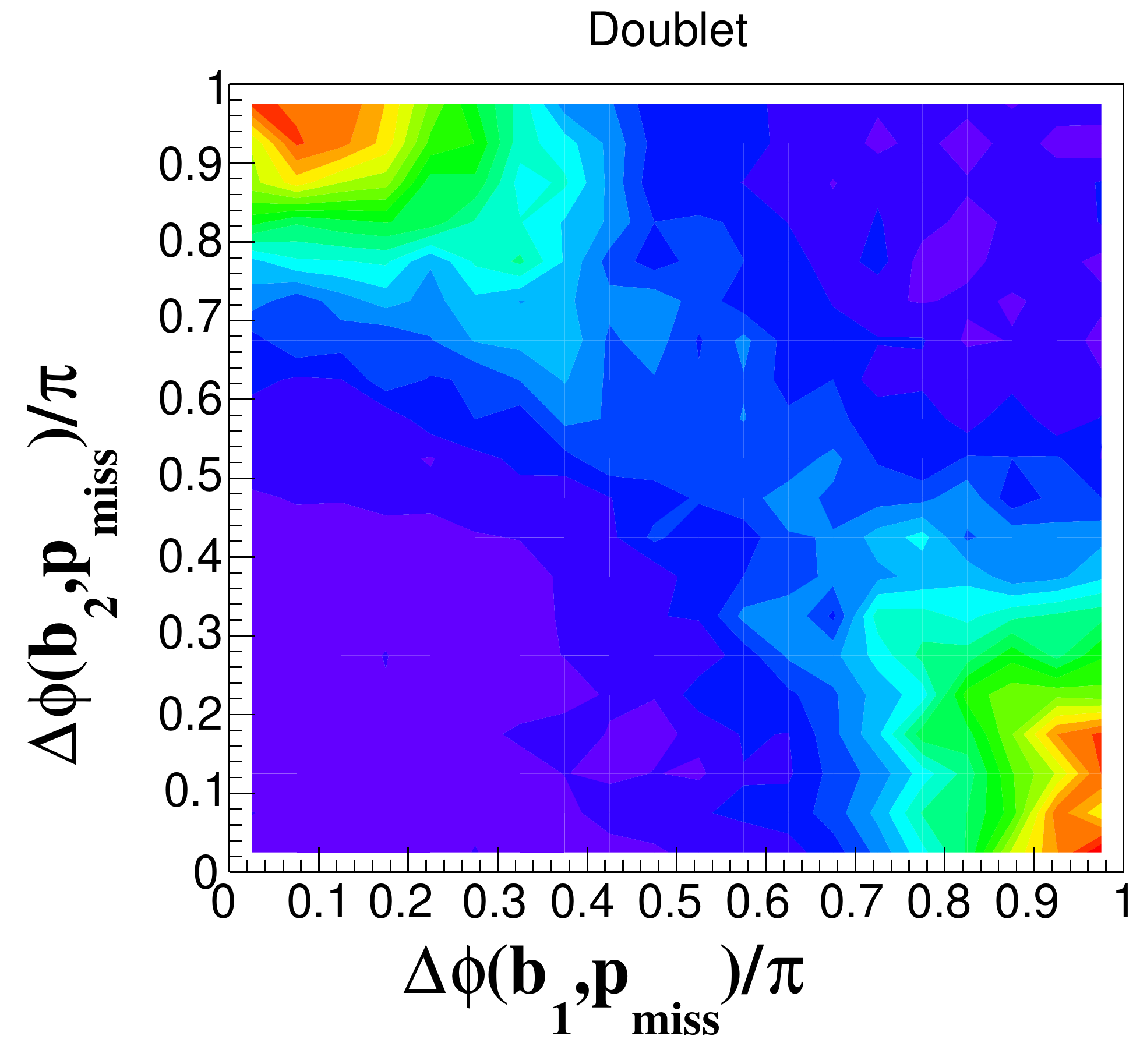}
	\includegraphics[width=0.3\textwidth]{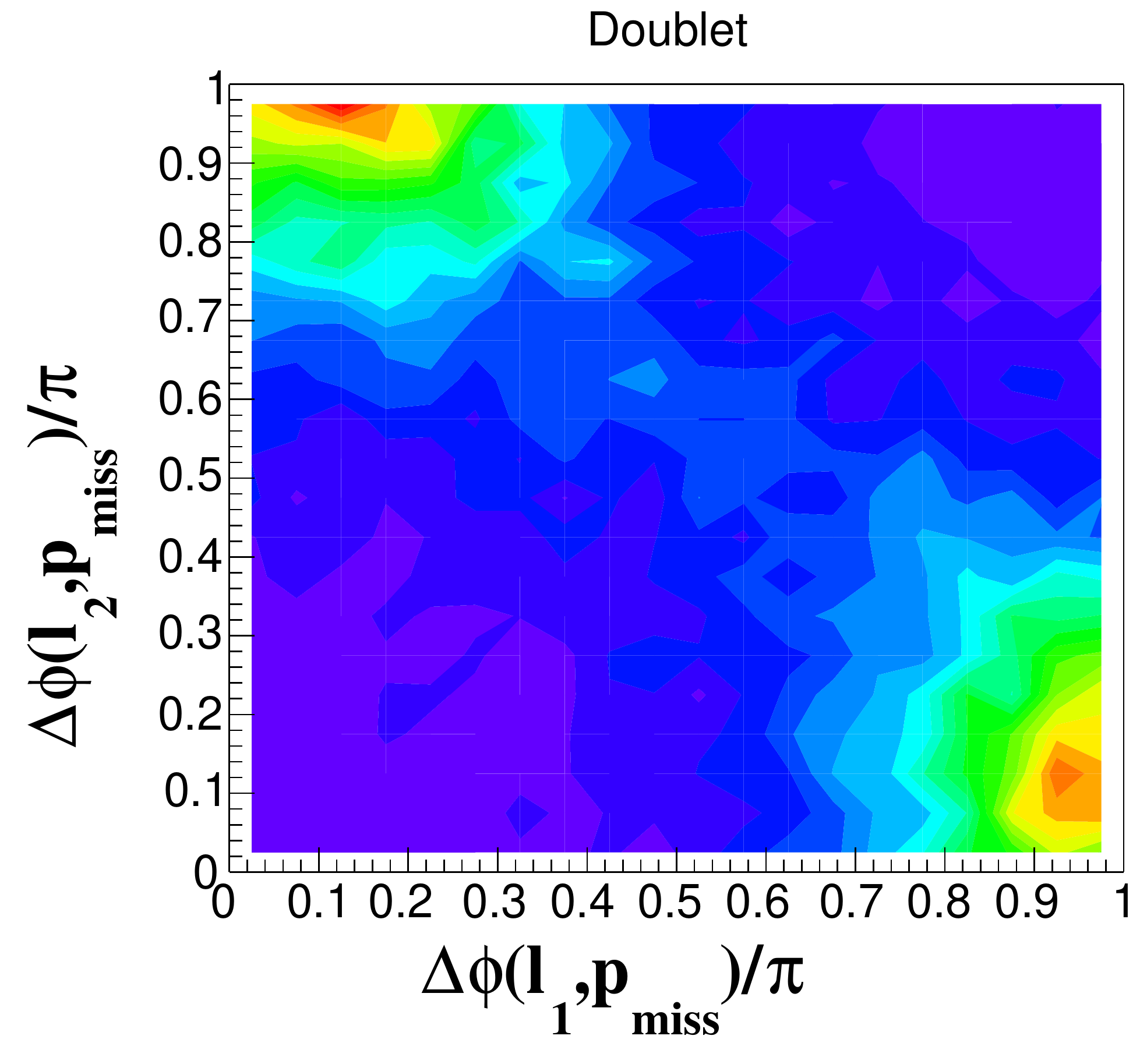}
	\includegraphics[width=0.3\textwidth]{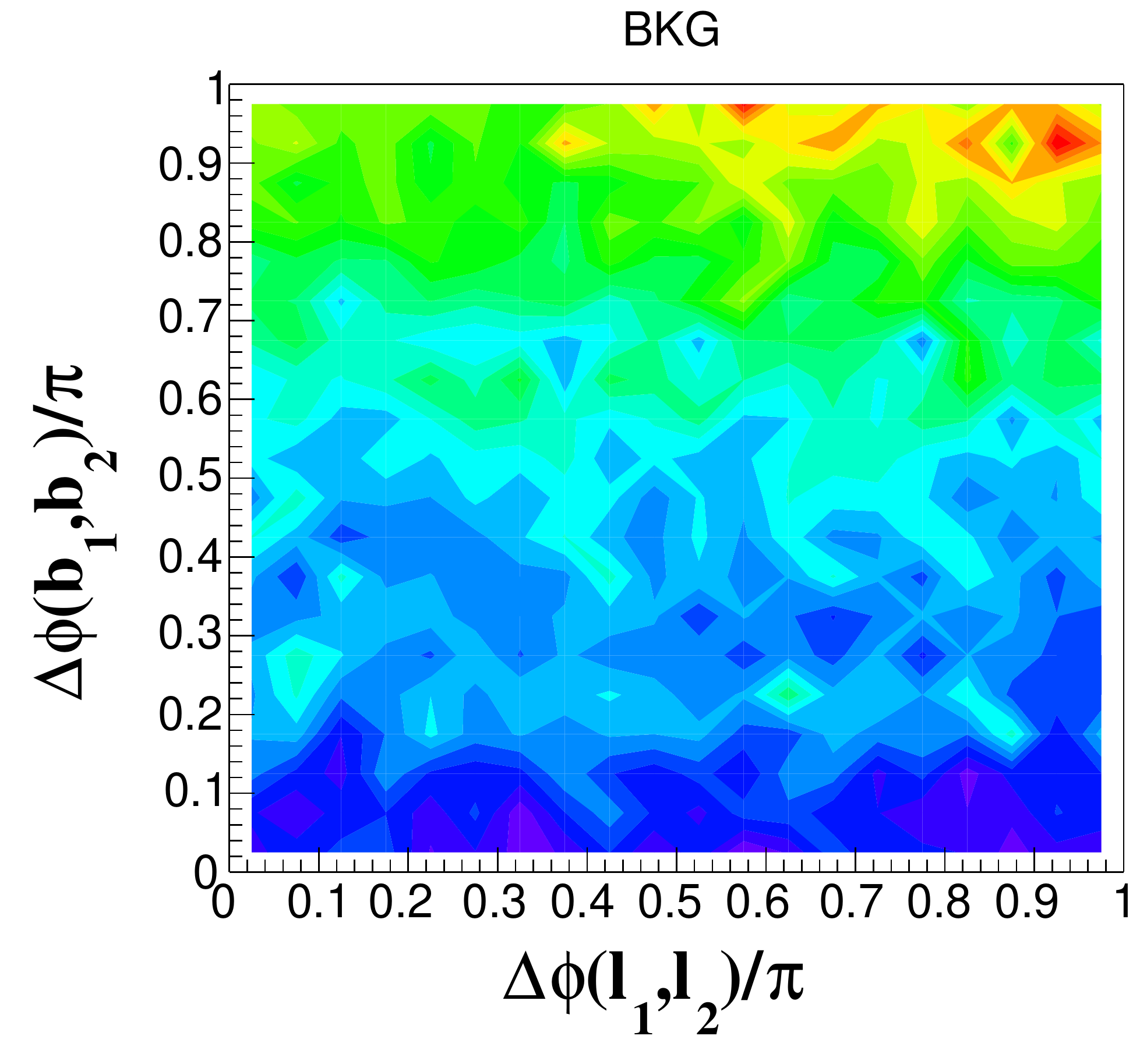}
	\includegraphics[width=0.3\textwidth]{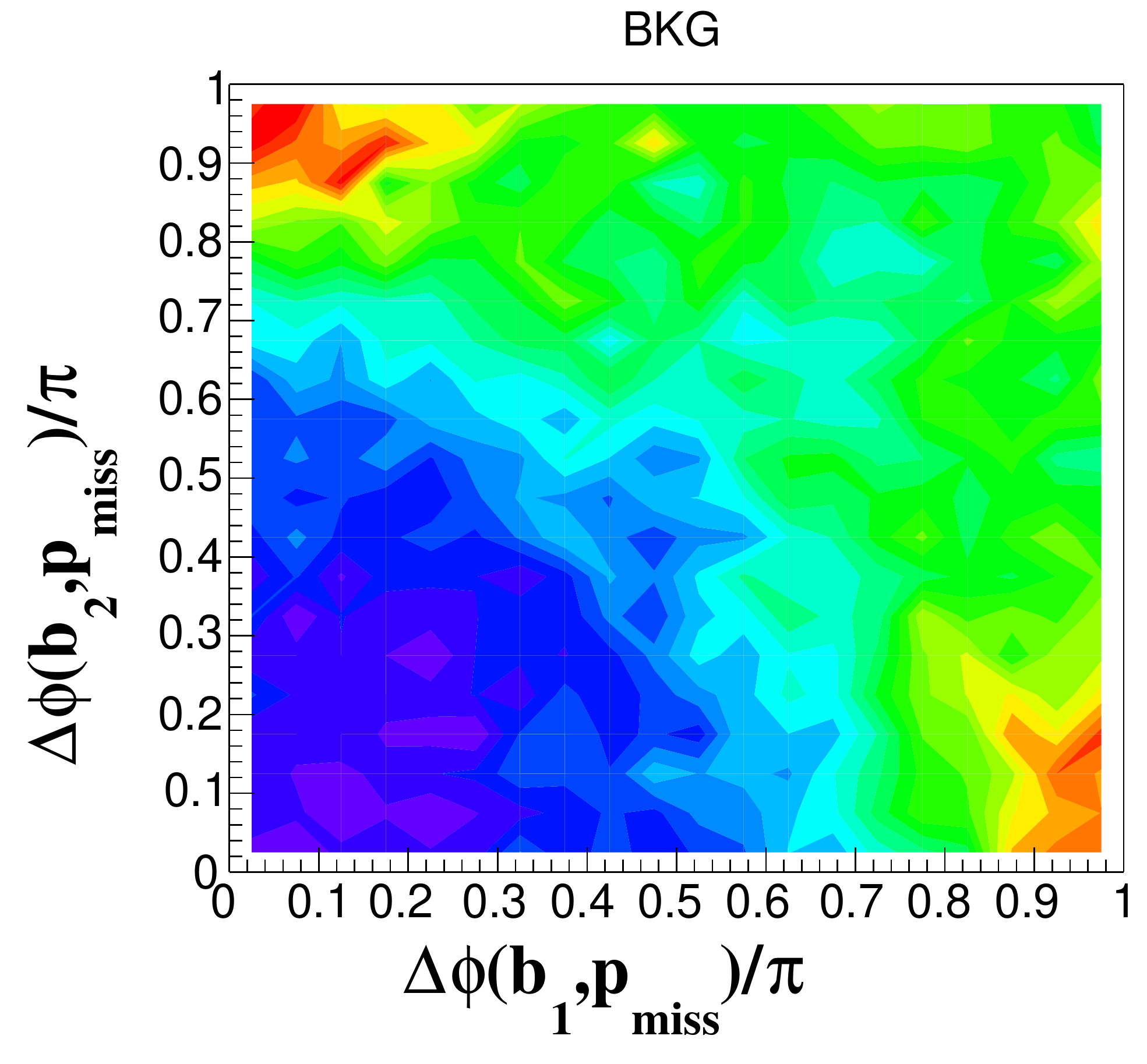}
	\includegraphics[width=0.3\textwidth]{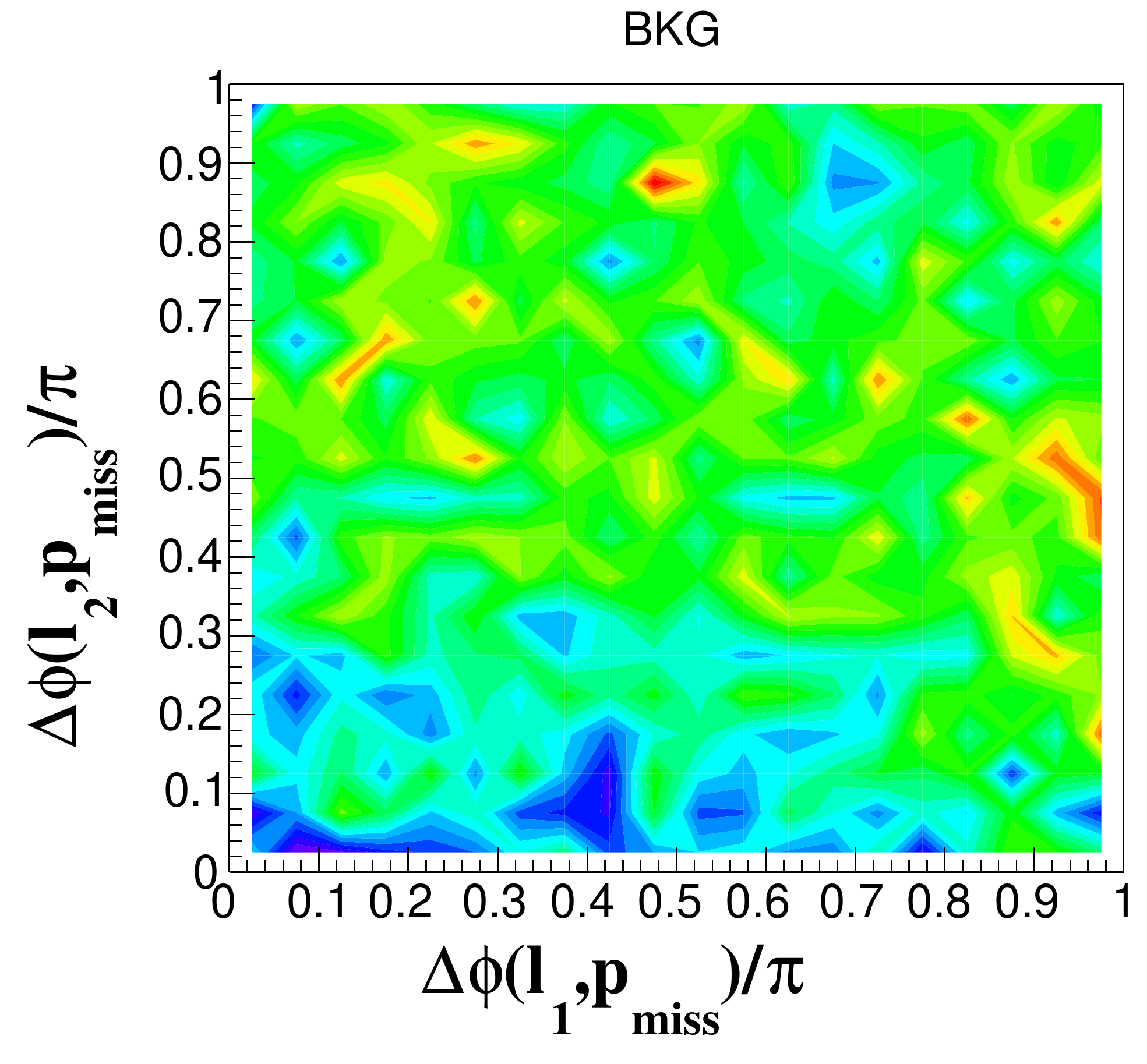}	
	\caption{	\label{fig:sig_bkg_sep_hist_2D} 
Normalized distributions for the two-dimensional correlations  between $\Delta\phi(l_1,l_2)$ vs. $\Delta\phi(b_1,b_2)$ (left-column), $\Delta\phi(b_1,p_{\text{miss}})$ vs. $\Delta\phi(b_2,p_{\text{miss}})$ (middle-column), and $\Delta\phi(l_1,p_{\text{miss}})$ vs. $\Delta\phi(l_2,p_{\text{miss}})$ (right-column)		
 for the signals and background at {\tt Delphes} level with selection cuts in Eq.~(\ref{eq:sel-cut}). The color description is the same as in  Fig.~\ref{fig:parton_2d}.}
\end{figure*}
\begin{figure}
	\centering
	\includegraphics[width=0.49\textwidth]{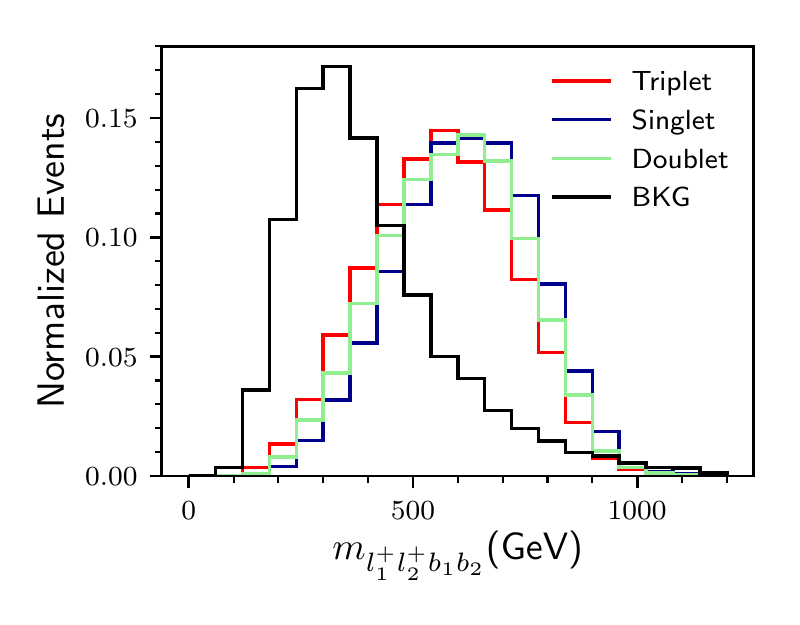}
	\caption{Normalized distributions for the  visible mass with selection cuts in Eq.~(\ref{eq:sel-cut}). For all signals, we choose $\lambda=0.003$ and $m_\phi = 1$ TeV.}
	\label{fig:Kinetic_dist}
\end{figure}
In this section, we discuss how to distinguish the three types of signals in the presence of background events, with a simplified detector-level simulation.
The backgrounds that can mimic our signal topology of two positively charged leptons ($2l^+$), two $b$-jets, and missing energy are~\cite{Modak:2022vjh,Ozsimsek:2021ivf,Hou:2020ciy,CMS:2017tec}
$t\bar{t}W^+$,  $t\bar{t}Z$, $t\bar{t}h$, $t\bar{t}W^+W^-$, $W^+ W^+ jj$, $ZZjj$,  $ZZW^+$, and $W^+W^-Z$.
In addition, the $t\bar{t}$+jets process, which has a large cross-section, can fake our signal if the lepton charges get misidentified~\cite{ATLAS:2018alq,ATLAS:2016kjm,Alvarez:2016nrz}.
We generate the signal ({\tt Singlet}, {\tt Doublet}, and {\tt Triplet}) and background events in {\tt MadGraph5\_aMC@NLO}v2.7.3~\cite{Alwall:2014hca} at leading order (LO) in QCD, without cuts on the final state particles, and with a dynamic choice of factorization scale given by $\sum_i M_i^T/2$, where $M_i^T$ is the transverse mass of the $i$-th final state particle. 
We use {\tt nn23lo1}~\cite{NNPDF:2014otw} for the PDFs. 
Background events are generated in {\tt MadGraph5} with the final states that can give rise to $2l^+2b+\cancel{E}_T$ at the detector level.
Events are then passed to {\tt PYTHIA8.2}~\cite{Sjostrand:2014zea} for showering and hadronization, followed by fast detector simulation in {\tt Delphes v3.4.2}~\cite{deFavereau:2013fsa}.
Events are selected at detector level with at least two $b$-tagged jets,  two positively charged leptons ($2l^+$), and missing transverse energy with the following selection level cuts 
\begin{eqnarray}\label{eq:sel-cut}
&p_T(l)>10~\text{GeV},~p_T(b)>20~\text{GeV},~\cancel{E}_T>10~\text{GeV},\nonumber\\
	&\Delta R (b,b) > 0.5,~\Delta R (b,l) > 0.4,~\Delta R (l,l) > 0.4,\nonumber\\
	&|\eta_b| < 2.5,~ |\eta_l| < 2.5.
\end{eqnarray}
With the selected background (BKG) and signal events, we study the one-dimensional and two-dimensional normalized distributions for the variables ($z_i$ and $\Delta\phi$ correlation, respectively) that we introduced in the previous section.
These are shown in Figs.~\ref{fig:sig_bkg_hist_1D}  ($z_i$) and \ref{fig:sig_bkg_sep_hist_2D} ($\Delta\phi$).
The qualitative features of the distributions at the detector level remain the same as the parton level distributions studied in the previous section. 
The $z_i$ distributions make clear distinctions between the {\tt Singlet} and {\tt Triplet} peaking in the right and left, respectively. 
Distributions are roughly symmetric around $z_i=0.5$ for the {\tt Doublet}. 
For the BKG, the $z_1$ distribution is symmetric around $z_1=0.5$ as well, while the $z_2$ distribution is asymmetric. 
In the case of all three signal benchmarks, the distributions for the  $z_1$ and $z_2$ are identical owing to the similar kinematics.	However, this not true for the background because the paired lepton and $b$ quark do not necessarily originate from a top quark. Additionally, $z_2$ peaks at lower values owing to the $p_T$ ordering of leptons. 
In the 2D $\Delta\phi$ distributions, the background events are distributed quite differently than the signals.
Therefore, it is possible to distinguish the signals based on these distributions, even with the detector-level events. 
 However, the total cross-section for the background is larger than that for the signal, and therefore we now discuss some kinematic cuts that are needed to reduce backgrounds and increase the signal sensitivity. 

We study various kinematic distributions, such as the transverse momenta ($p_T$) of the final state particles, total hadronic energy ($H_T$), and visible mass ($m_{l_1^+l_2^+b_1b_2}$). 
We find that visible mass, the normalized distribution shown in Fig.~\ref{fig:Kinetic_dist}, can be used to suppress the background and enhance the significance of all three signals. 
In particular, a cut on the visible mass of
\begin{equation}\label{eq:mvis-cut}
	m_{l_1^+l_2^+b_1b_2}>430~~\text{GeV}
\end{equation}
maximizes the signal significance.
The signal and background processes are summarized in Table~\ref{tab:Signal-bkg}, showing their cross sections at the generation level and the expected number of events for an integrated luminosity of ${\cal L}=300$ fb$^{-1}$ after the visible mass cut. 
The total number of background events is expected to be $208$  in the $2l^+ 2b \cancel{E}_T$ final state. The signal significance expected for the three signals are given in Table~\ref{tab:significance} for integrated luminosities of ${\cal L}=150$ fb$^{-1}$ (currently available), $300$ fb$^{-1}$ (next phase of LHC), and $3000$ fb$^{-1}$ (high luminosity phase).
We calculate the signal significance~\cite{Cowan:2010js} using
\begin{equation}
	S = \sqrt{2\left[(s+b) \log\left(1+\frac{s}{b}\right)-s\right]},
\end{equation}
where $s$ and $b$ stand for the total number of signal and background events surviving after cuts.
For the chosen coupling of $\lambda=0.003$ and $m_\phi=1$ TeV, the significance for the {\tt Triplet} and {\tt Doublet} signals are higher than the {\tt Singlet} signal. 
A luminosity of about $1300$ fb$^{-1}$ is required for discovery of {\tt Singlet}, whereas the {\tt Doublet} and {\tt Triplet} can be discovered with about $790$ fb$^{-1}$ and $930$ fb$^{-1}$ of luminosity, respectively  with $5\sigma$ significance.

\begin{table*}\caption{\label{tab:Signal-bkg} The signal and the background cross sections and the expected number of events for an integrated luminosity of ${\cal L}=300$ fb$^{-1}$ with selection cuts in Eq.~(\ref{eq:sel-cut}) as well as the visible mass cut in Eq.~(\ref{eq:mvis-cut}). The contents in the parenthesis of the first column correspond to the final states contributing to the $2l^+2b\cancel{E}_T$ final state. For all signals, we choose $\lambda=0.003$ and $m_\phi = 1$ TeV.}
	\renewcommand{\arraystretch}{1.50}
	\centering	
	\begin{tabular*}{\textwidth}{@{\extracolsep{\fill}}lccc@{}}\hline		
		Process (generated up-to) &  Cross section (fb) & Efficiency ($\epsilon$) & \begin{tabular}{c}
			Expected Events \\ ($2l^+ 2b \cancel{E}_T$)
		\end{tabular}  \\	\hline 
		{\tt Singlet}: $tt~(2l^+2b\cancel{E}_T)$ &  $1.121 $  & $10.62~\% $ & $35.7 $ \\	\hline 
		{\tt Doublet}: $tt~(2l^+2b\cancel{E}_T)$ & $1.348 $ &$11.40~\% $ & $46.1 $ \\	\hline 
		{\tt Triplet}: $tt~(2l^+2b\cancel{E}_T)$ & $1.166 $ &$12.06~\% $ & $42.2$ \\	\hline \hline
		{\tt B1:} $t\bar{t}W^+~(2l^+2b2j\cancel{E}_T)$ & $ 9.2$ &$ 3.75~\%$ & $103.5$ \\	\hline 
		{\tt B2:} $t\bar{t}Z~(t\to l^+b\cancel{E}_T,\bar{t}\to \bar{b}jj+\bar{b}l^-\cancel{E}_T,Z\to 2l)$ &  $8.056 $ &$1.76~\% $ & $42.6 $  \\	\hline 
		{\tt B3:} $t\bar{t}+jj~(t/\bar{t}\to l^+/l^-)$ & $29482.0 $ &$4.7\times 10^{-4}~\% $ & $41.6 $ \\	\hline 	
		{\tt B4:} $t\bar{t}h+jj~(t/\bar{t}\to l^+/l^-,h\to\text{all})$ & $23.68 $ &$0.14~\% $ & $10.3 $ \\	\hline 					
		{\tt B5:} $t\bar{t}W^+W^-$~($t/W^+\to l^+,~\bar{t}/W^-\to all$)    & $0.398 $ &$4.94~\% $ & $5.9 $ \\	\hline 
		{\tt B6:} $W^+ W^+ jj~(2l^+2j\cancel{E}_T)$  & $8.967 $ &$0.11~\% $ & $2.9$ \\	\hline 
		{\tt B7:} $ZZjj~(4l2j)$ & $12.69 $ &$0.02~\% $ & $0.8 $ \\	\hline 
		{\tt B8:} $ZZW^+~(4ljj+3ljj\cancel{E}_T)$  & $ 0.4$ &$0.17~\% $ & $0.2 $ \\	\hline 
		{\tt B9:} $W^+W^-Z~(3l2j\cancel{E}_T)$ & $1.61 $ &$\le 10^{-3}~\% $ & $0 $ \\	\hline

		Total Background & &&$207.8$ \\	\hline 
	\end{tabular*}
\end{table*}
\begin{table}\caption{\label{tab:significance} Signal significance using only the total number of events with integrated luminosities of ${\cal L}=150$ fb$^{-1}$, $300$ fb$^{-1}$, and $3000$ fb$^{-1}$ with selection cuts in Eq.~(\ref{eq:sel-cut}) as well as the visible mass cut in Eq.~(\ref{eq:mvis-cut}). For all the signals, we choose $\lambda=0.003$ and $m_\phi = 1$ TeV.}
	\renewcommand{\arraystretch}{1.50}
	\centering	
	\begin{tabular*}{0.46\textwidth}{@{\extracolsep{\fill}}lcccc@{}}\hline		
		Signal  &  $150$ fb$^{-1}$ &  $300$ fb$^{-1}$ &  $ 3000$ fb$^{-1}$ & ${\cal L}$ for $5\sigma$ C.L.\\ \hline
		{\tt Singlet} &  $1.70 $  &    $2.40 $  &   $7.59 $  &  $\simeq 1300$ fb$^{-1}$\\ \hline
		{\tt Doublet} &  $2.18  $  &   $3.09 $ &   $9.77  $  &  $\simeq 790$  fb$^{-1}$ \\ \hline
		{\tt Triplet} &  $2.01 $   &   $2.84 $ &   $8.97  $  &   $\simeq 930$ fb$^{-1}$\\ \hline
	\end{tabular*}	
\end{table}
\begin{table}\caption{\label{tab:asym_sig_bkg} Values of asymmetries for the signals along with the backgrounds with selection cuts in Eq.~(\ref{eq:sel-cut}) as well as the visible mass cut in Eq.~(\ref{eq:mvis-cut}). For all the signals, we choose $\lambda=0.003$ and $m_\phi = 1$ TeV.}
	\renewcommand{\arraystretch}{1.50}
	\centering	
	\begin{tabular*}{0.46\textwidth}{@{\extracolsep{\fill}}lccccc@{}}\hline		
		& $\mathcal{A}_{z_1}$ & $\mathcal{A}_{z_2}$ & $\mathcal{A}_{lb}$ & $\mathcal{A}_{ll}$ & $\mathcal{A}_{bb}$ \\ \hline
		{\tt Singlet} & $0.52$ & $0.49$  &$0.28 $ &$0.35 $ &$-0.04 $ \\ \hline 
		{\tt Doublet} & $0.13$&$0.09  $&$-0.12  $&$ -0.03$ & $0.08$ \\ \hline 	
		{\tt Triplet} &$ -0.27$&$-0.32 $&$-0.36$  &$-0.09$  &$ 0.32$ \\ \hline 
		BKG &$ 0.14$&$ -0.28 $ &$-0.31 $ &$ -0.50$ & $-0.25$ \\ \hline 			
		{\tt Singlet} + BKG &$ 0.19$ &$ -0.16 $ &$-0.22 $ &$ -0.38$  &$-0.21 $ \\ \hline 
		{\tt Doublet} + BKG &$ 0.14$&$-0.21 $ &$-0.27 $ &$ -0.42$ &$ -0.19 $\\ \hline 		
		{\tt Triplet} + BKG & $0.07$&$ -0.28 $&$-0.31 $ &$-0.43$  &$ -0.15$ \\ \hline 
	\end{tabular*}
\end{table}
\begin{table}\caption{\label{tab:diff-signals} Luminosity (${\cal L}$) required to distinguish between the models at $2\sigma$, $3\sigma$, and $5\sigma$ C.L. based on the asymmetries with selection cuts in Eq.~(\ref{eq:sel-cut}) as well as the visible mass cut in Eq.~(\ref{eq:mvis-cut}). We choose $\lambda=0.003$ and $m_\phi = 1$ TeV for all the signals. }
	\renewcommand{\arraystretch}{1.50}
	\centering	
	\begin{tabular*}{0.46\textwidth}{@{\extracolsep{\fill}}lccc@{}}\hline		
		&\multicolumn{3}{c}{Luminosity required ( fb$^{-1}$)} \\ \hline
		Signal\_1 vs. Signal\_2 &  $2\sigma$ C.L.& $3\sigma$ C.L. & $5\sigma$ C.L. \\ \hline
		{\tt Singlet} vs. {\tt Doublet}&$455$ & $1023 $  & $2842$ \\ \hline
		{\tt Singlet} vs. {\tt Triplet}& $100$& $226 $ & $628$  \\ \hline
		{\tt Doublet} vs. {\tt Triplet}&$330$ & $742 $ 	& $2061$  \\ \hline
	\end{tabular*}
\end{table}
Having discussed the discovery potential for the signals, we now turn to our primary goal of this analysis, which is to distinguish among the three signals with the help of the observables discussed in Sec.~\ref{sec:obs}.
To this end,  we first calculate the asymmetries for all the variables and summarize them in Table~\ref{tab:asym_sig_bkg} for all three signals and backgrounds separately, as well as, signals in the presence of backgrounds. 
The numbers in the first three rows show that all three signals can be identified by looking at the value and the sign of the five asymmetries. 
However, these asymmetries are affected by the background events, as shown in the last three rows.
We calculate the differences between any two signals in the presence of backgrounds using all five asymmetries, given in Table~\ref{tab:asym_sig_bkg}, in terms of the $\chi^2$ function
\begin{equation}\label{eq:chi2}
\chi^2=\sum_{i} \left| \frac{ {\cal A}_i(\text{Signal\_$1$+BKG})- {\cal A}_i(\text{Signal\_$2$+BKG})}
{\delta {\cal A}_i(\text{BKG})} \right|^2.
\end{equation}
Here $\delta {\cal A}_i(\text{BKG})=\sqrt{ \frac{1-{\cal A}_i^2(\text{BKG})}{{\cal L}\sigma(\text{BKG})}  }$ is the statistical uncertainty due to the SM backgrounds with $\sigma(\text{BKG})$ being the total background cross section. 
Signal\_$j$ ($j=1,2$) denote any two among the {\tt Singlet}, {\tt Doublet}, and {\tt Triplet} signals.
We estimate the luminosity required for $2\sigma$, $3\sigma$, and $5\sigma$ C.L. separability among the signals using the above $\chi^2$ functions and quote them in Table~\ref{tab:diff-signals}. 
The {\tt Singlet} and the {\tt Triplet} are quite different since the former (latter) decays only into right-(left-) chiral top quarks; a luminosity of about $628$ fb$^{-1}$ is required to achieve separability at $5\sigma$ C.L., although neither of them can be discovered for that luminosity with the expected number of events, see Table~\ref{tab:significance}. 
On the other hand, we require about $2842$ fb$^{-1}$ and $2061$ fb$^{-1}$   of luminosity for $5\sigma$ separation between the {\tt Singlet} versus {\tt Doublet} and {\tt Doublet} versus {\tt Triplet}, respectively, as the two top quarks have both chiralities originating from the {\tt Doublet}. 
Thus, to make distinctions between the signals involving the {\tt Doublet}, we need higher luminosity than what is required to discover them.

\section{Conclusion}\label{sec:conclusion}
In this work, we have discussed how to measure the electroweak quantum numbers of BSM color sextet scalar and vector particles. 
While all the sextet particles that we consider decay into a {\it like}-sign top quark pair, the top quarks have identical chirality for the two sextet scalars and opposite chirality for the sextet vector. 
Furthermore, one of the scalars give rise to left-handed top quarks while the other decays to right-handed ones.
These features can be captured by several kinematic variables that rely only on visible final states.
One such variable is the well known visible energy fraction ($z_i$) of final state leptons,  Eq.~(\ref{eq:energy-frac}). 
Here we construct three additional variables, defined in Eq.~(\ref{eq:2d_asym}), that depend on the angular correlation between the final state leptons and $b$-jets. 
All of these variables are sensitive to the polarization of the top quarks, and in a combined fashion can distinguish among the three possible sextet states.

Through a parton level analysis, we first demonstrate the utility of the visible energy fraction variables and the angular correlation variables.
We study their distributions in Figs.~\ref{fig:parton_1d} and~\ref{fig:parton_2d}, and compute a set of asymmetries in Table~\ref{tab:asymmetris-parton} to identify the differences among the three signals. 
These show that the asymmetries can fully distinguish, as well as, identify the quantum numbers of the sextet states, in principle.

We then implement a simplified detector level simulation, taking into account possible SM backgrounds, to verify how well we can differentiate among the three types of signals. 
We find that while the three signals can be distinguished among themselves even at the detector level, the inclusion of SM background reduces the difference among the signals.
Nonetheless, with sufficient statistics within the reach of the high-luminosity phase of the LHC, the three types of signals can still be distinguished, as demonstrated in Table~\ref{tab:diff-signals}. 
We find higher luminosities are required to make all these distinctions than the luminosities required to discover them.
To summarize, if discovered at the LHC, it is possible to measure the electroweak quantum numbers of BSM color sextet particles using top quark polarization and spin correlation observables.
\section*{Acknowledgements}
S.K. is supported in part by the U.S. National Science Foundation (NSF) grant PHY-1915314 and the DOE contract DE-AC02-05CH11231. S.K. thanks IISER Kolkata for hospitality during various stages of this work.
R.R. would like to acknowledge support from the Department of Atomic Energy, Government of India, for the Regional Centre for Accelerator-based Particle Physics (RECAPP), Harish Chandra Research Institute.

\bibliographystyle{utphysM}
\bibliography{refence,ReferencesttZ}

\end{document}